\documentclass[12pt, preprint]{aastex}
\usepackage{lineno}
\usepackage{epsfig, longtable}

\shorttitle{\emph{Fermi} LAT view of MAGN }
\shortauthors{Abdo et al.}

\title{\emph{Fermi} Large Area Telescope Observations of Misaligned AGN}

\author{
A.~A.~Abdo\altaffilmark{1,2}, 
M.~Ackermann\altaffilmark{3}, 
M.~Ajello\altaffilmark{3}, 
L.~Baldini\altaffilmark{4}, 
J.~Ballet\altaffilmark{5}, 
G.~Barbiellini\altaffilmark{6,7}, 
D.~Bastieri\altaffilmark{8,9}, 
K.~Bechtol\altaffilmark{3}, 
R.~Bellazzini\altaffilmark{4}, 
B.~Berenji\altaffilmark{3}
% ...
}
\altaffiltext{1}{Space Science Division, Naval Research Laboratory, Washington, DC 20375, USA}
\altaffiltext{2}{National Research Council Research Associate, National Academy of Sciences, Washington, DC 20001, USA}
\altaffiltext{3}{W. W. Hansen Experimental Physics Laboratory, Kavli Institute for Particle Astrophysics and Cosmology, Department of Physics and SLAC National Accelerator Laboratory, Stanford University, Stanford, CA 94305, USA}
\altaffiltext{4}{Istituto Nazionale di Fisica Nucleare, Sezione di Pisa, I-56127 Pisa, Italy}
\altaffiltext{5}{Laboratoire AIM, CEA-IRFU/CNRS/Universit\'e Paris Diderot, Service d'Astrophysique, CEA Saclay, 91191 Gif sur Yvette, France}
\altaffiltext{6}{Istituto Nazionale di Fisica Nucleare, Sezione di Trieste, I-34127 Trieste, Italy}
\altaffiltext{7}{Dipartimento di Fisica, Universit\`a di Trieste, I-34127 Trieste, Italy}
\altaffiltext{8}{Istituto Nazionale di Fisica Nucleare, Sezione di Padova, I-35131 Padova, Italy}
\altaffiltext{9}{Dipartimento di Fisica ``G. Galilei", Universit\`a di Padova, I-35131 Padova, Italy}
% ...

\begin{abstract}

Analysis is presented on 15 months of data taken with the Large Area Telescope (LAT) 
on the \emph{Fermi} Gamma-ray Space Telescope for 11 non-blazar AGNs, including
7 FRI radio galaxies and 4 FRII radio sources consisting of 2 
FRII radio galaxies and 2 steep spectrum radio quasars. The broad line 
FRI radio galaxy 3C 120 is reported here as a $\gamma$-ray source for the first time. 
The analysis is based on directional associations of LAT sources with 
radio sources in the 3CR, 3CRR and MS4 (collectively referred to as 3C-MS) 
catalogs.  
Seven  of the eleven LAT sources associated with 3C-MS radio sources have spectral 
indices larger  than 2.3 and, except for the FRI radio galaxy NGC 1275 that shows
possible spectral curvature, are well described by a power law.
No evidence for time variability is found for any sources other than NGC 1275.
The $\gamma$-ray luminosities of FRI radio galaxies are 
significantly smaller than those of  BL Lac 
objects detected by the LAT, whereas the $\gamma$-ray luminosities of FRII sources 
are quite similar to 
those of FSRQs, which could
reflect different beaming factors for the $\gamma$-ray emission. 
A core dominance study of the  3CRR sample indicate that sources closer to the jet axis are preferentially 
detected with the {\it Fermi}-LAT, insofar as the $\gamma$-ray--detected misaligned AGNs 
have larger core dominance at a given average radio flux. The results are discussed in 
view of the AGN unification scenario. 

\noindent {\bf Contact authors: P. Grandi, G. Malaguti, G. Tosti, C. Monte}
\end{abstract}


\begin{document}



\section{Introduction }

The unification scenario (e.g., Urry \& Padovani 1995) for active galactic nuclei (AGNs) 
explains a large variety of AGN properties
in terms of viewing angle towards a system consisting of an obscuring
torus, an accretion disk providing fuel for a supermassive black hole, and broad- and narrow-line emission 
regions surrounding the black hole. 
Left unexplained is the dichotomy between radio-quiet and radio-loud AGNs. 
The latter objects contain jets of collimated plasma  
ejected with relativistic speeds transverse to the plane of the accretion disk. 
Such jets are very weak or absent in radio-quiet AGNs.

For the radio-loud AGNs, there are two main causes of anisotropy: 
the obscuring material of the torus that is roughly coplanar with and 
probably feeds the accretion disk,  and  the radio-emitting jets.

Decelerating  jets  and kpc scale edge-darkened lobes are found in the 
weaker FRI radio galaxies,  while relativistic jets and edge-brightened radio lobes are found
in the stronger FRII radio galaxies (Fanaroff \& Riley 1974).  According to the unification scenario, BL Lac objects and 
Flat Spectrum Radio Quasars (FSRQ) represent FRI and FRII radio galaxies, respectively, viewed
nearly along the jet axis. In these sources, non-thermal radiation emitted
from jets
at the pc scale is  
amplified by relativistic effects to produce flat radio spectrum sources with large optical polarization and 
strong optical variability, which furthermore are often found to exhibit superluminal motion in 
detailed radio monitoring. 
Sources with these attributes are 
collectively referred to as blazars. Multiwavelength observations show that blazars 
typically exhibit two-peaked broad-band
spectral energy distributions (SEDs) from radio to $\gamma$ rays with the 
non-thermal electron-synchrotron radiation forming the lower-energy radio to X-ray 
emission, and 
Compton processes likely making the $\gamma$ radiation (see, e.g., B\"ottcher 2007, for review). 
Due to  strong Doppler boosting,  $\gamma$-ray blazars  are detected from 
redshifts as large as $z \approx 3$ and with apparent $\gamma$-ray luminosities
sometimes exceeding $\approx 10^{49}$ erg s$^{-1}$. 
The largest identified source population in the $\gamma$-ray sky are blazars 
(Hartman et al.\ 1999; Abdo et al.\ 2010a,b).

By comparison, misaligned AGNs (MAGNs), with jets pointed away from the observer,  are not 
favored GeV sources. By MAGN we mean radio-loud AGNs with misdirected  jets that
display steep radio spectra ($\alpha_{r} \ge 0.5$ with the usual convention that 
$F_\nu \propto \nu^{-\alpha_r}$) and bipolar or quasi-symmetrical structures in radio maps. 
The larger jet inclination angle, in contrast with blazars, deboosts the radiation to 
make the relativistic jet radiation weaker than other potential sources of radio emission
in these objects, such as synchrotron radiation from mildly relativistic outflows or extended 
radio lobe emission.  The high-significance detections of NGC 1275 (Abdo et al.\ 2009a) 
and the Centaurus A radio galaxy confirming 
the EGRET result (Sreekumar et al.\ 1999; Abdo et al.\ 2010c,d) 
in only three months of scientific 
observations with the \emph{Fermi} LAT
(Atwood et al.\ 2009a) clearly shows
that misaligned radio-loud AGNs are another and
potentially very interesting class of  gamma-ray emitters.  
Indeed several  other MAGNs are listed in the first-year LAT AGN Catalog  paper 
(1LAC; Abdo et al.\ 2010a). Very recently a LAT  discovery of very high energy emission ($>100$  GeV) from the 
radio galaxy  IC~310 has been reported (Neronov et al. 2010).  This head-tail radio source, situated in the Perseus cluster,  could belong to a new MAGN class of  GeV emitters.
The observed high energy  photons could not originate in a jet but  be produced   at the bow shock 
formed in interaction of fast motion of the galaxy through the dense intercluster medium.

In this paper, we present a dedicated study  of the $\gamma$-ray  properties of
all radio sources belonging to the Cambridge (3CR and 3CRR) and
the Molonglo (MS4) catalogs associated with LAT detections in the first 15 months of sky survey, 
as described in Section 2.
Our sample consists of  eleven radio sources,  
all of which were reported in the 1LAC except for the radio galaxy  3C~120.
LAT observations and analysis are described in Section 3, with results of the analysis
given in Section 4. The properties of the MAGNs and Fermi LAT blazars are compared in 
Section 5, and the implications are discussed in Section 6. We conclude in Section 7.
 
In the following, we use a $\Lambda$CDM cosmology with values within $1\sigma$ of
the WMAP results (Komatsu et al.\ 2008); in particular, we use $h = 0.71$,
$\Omega_m = 0.27$, and $\Omega_\Lambda = 0.73$, where the Hubble constant
$H_0=100h$ km s$^{-1}$ Mpc$^{-1}$.

\section{The Sample }

Our sample includes 11 MAGNs made up of  nine radio galaxies, consisting of 7 FRI and 2 FRII radio galaxies,
 and two  steep spectrum radio quasars (SSRQs) with radio luminosities comparable to FRII sources. 
One of the two SSRQs, 3C~380,  is sometimes  classified as a compact steep spectrum (CSS) source, 
since most of its 5 GHz flux is emitted at galactic scales rather than at hundreds of kpc,
 as usually observed in FRII quasars (Wilkinson et al.\ 1991). In the  unification scenario proposed for Radio Loud AGNs, 
FRI and FRII radio galaxies are the parent population of BL Lac objects and FSRQs, respectively.
The larger number, seven, of FRI sources compared to the four FRII objects may already reflect an interesting feature
of the emission process insofar as BL Lacs and FSRQs are about equally represented in the 1LAC 
(Abdo et al.\ 2010a).

With the exception of 3C 120, all of these sources are already reported in the 1LAC  paper and
have been established as associations by cross-correlating the 1-year catalog sources 
with the 3CR catalog (Bennett 1962; Spinrad et al.\ 1985),  the revised 3CRR catalog (Laing et al.\ 1983), 
and the Molonglo Southern 4 Jy Sample (MS4: Burgess \& Hunstead 2006ab). 
These surveys are flux-limited, with the 3CR, 3CRR, and MS4 flux limits of 
9 Jy at 178 MHz, $10.9$ Jy---also at 178 MHz, and $4$ Jy at 408 MHz, respectively.
The low-frequency  selection criterion
favors the detection of  radio sources characterized by steep-spectrum synchrotron emission from 
extended lobes. Thus the use of these catalogs (jointly referred to as the 3C-MS catalogs) 
 for associations would preferentially select radio sources with large viewing angles. The 3C and 3CRR catalogs 
cover most part of the northern sky, with declination Dec $> -5^{\circ}$ and Dec $\ge 10^\circ$, respectively, while
the MS4 catalog covers most of the southern sky, with $-85^{\circ}<$Dec$<-30^{\circ}$.
Finally, both optical and radio classifications are available for the majority of the  sources.  In particular,  the 3CRR objects have been extensively studied from radio to X-rays.

LAT sources are positionally associated at high probability with the seven FRI radio galaxies and the two SSRQs,
and are included in the 1LAC (Table 1 in Abdo et al.\ 2010a).  
A LAT source is characterized by a high association probability (P=87$\%$) with the  FRII radio galaxy, 3C~111, 
but being  located at low Galactic latitude, is not technically part of the 1LAC which is restricted to $|b|>10^\circ$.
It is, however, reported in Table 2 of  the 1LAC. 
The other FRII radio galaxy, PKS~0943$-$76, is  within the 95$\%$ error circle radius of  1FGL~J0940.2$-$7605 and 
is considered a plausible association with the LAT $\gamma$-ray source, and 
is listed among the AGN affiliations in Table 3 of the 1LAC paper. 
As its LAT association is less secure,  we analyze the data of this radio galaxy 
keeping in mind its less secure association.
Finally we note that 3C~207 shows  a very high association  probability (P=99\%)  with  1FGL~J0840.8$+$1310.
Two other AGNs with association probabilities (51\% and  71\%) lower than for 3C 207 are found within the $95\%$ error radius.
3C~120 does not appear in the 1LAC, so we performed a new analysis now including all the available 15-month LAT data.

Table 1 gives the 1FGL LAT source name, probable associations including their RA and Dec\footnote{Coordinates from NASA/IPAC   EXTRAGALACTIC   DATABASE: NED}, redshift,  radio 
(FRI, FRII, or CSS),  optical (G - galaxy, BLRG - Broad Line Radio Galaxy) and radio-optical (SSRQ - Steep Spectrum Radio Quasar) type of the MAGN sources.  
Also reported is the 
radio core dominance (CD) at 5 GHz.  The CD value, considered a good indicator of the jet orientation, is defined as
$CD=Log(S_{\rm core}/[S_{\rm tot}-S_{\rm core}])$, where $S_{\rm core/tot}$ is the core/total flux density 
referred to the source rest frame (Scheuer \& Readhead 1979).
If  we assume that the intrinsic power of the core is a fixed fraction $f$ 
of the power of the extended components, CD is proportional to the beaming ($CD=f \delta^{3+\alpha}$). 
If the bulk Lorentz factor ($\gamma$) of the emitting plasma is similar  for  
all the sources with the same radio morphology,  CD can be directly related to 
inclination angle $\theta$.  The CD parameter suffers, however,
 from several systematic uncertainties, for example, core variability,  and radio map
resolution and sensitivity.  Nevertheless,  it provides a quantitative measure
to directly connect observational quantities to AGN properties, in particular, beaming and inclination angle.

 For the MAGNs quoted in Morganti et al.\ (1993) and Burgess \& Hunstead (2006ab), 
the reported 5 GHz core fluxes density are K-corrected using 
$\alpha_{core}$=0, while for the extended  component the correction has been performed using the spectral index  listed in those papers.
 For all the other 3CRR objects in the sample, CD is calculated  using the total 178 MHz flux, the 5 GHz core flux and $\alpha_r$ between 178 and 750 MHz  provided by the  on-line 3CRR database.\footnote{http://3crr.extragalactic.info/cgi/database}
The  total 178 MHz flux density  was converted into the total 5~GHz value following  the method proposed by Fan $\&$ Zhang (2003).
This approach made it possible to successively extend the CD study to the entire 3CRR sample (see Section 5).


\section{LAT observations  and Data Analysis}\label{sect:LAT_data_analysis}

The {\it Fermi}-LAT is a pair-conversion $\gamma$-ray telescope sensitive to  photon energies from 20~MeV~ to $>300$~GeV.  The LAT has a large peak effective area ($\sim 8000$ cm$^2$ for 1 GeV~ photons in the event class considered here), viewing $\sim$2.4 sr of the full sky with excellent angular resolution (68\% containment radius better than 
$\sim1^{\circ}$ at $E = 1$ GeV). It operates mainly in sky-survey mode, observing the entire sky every 3 hours. 
For a detailed description of the LAT, see Atwood et al.\ (2009).

We analyzed LAT data collected during the first 15 months of operation, from
August 4, 2008  to November 8, 2009. We kept only events in the ``diffuse" class, 
with energies in the range 0.1 -- 100 GeV, and with reconstructed zenith angle $<105^{\circ}$ in order 
to reduce the bright $\gamma$-ray albedo from the Earth. Also, we excluded the time intervals when the rocking angle was more than 52$^{\circ}$ and when the {\it Fermi} satellite was within the South Atlantic Anomaly.
The standard {\it Fermi-LAT ScienceTools} software package\footnote{http://fermi.gsfc.nasa.gov/ssc/data/analysis/documentation/Cicerone/} (version v9r15p5 ) was used with the P6$\_$V3  set of instrument response functions.  The spectral study was performed using the unbinned maximum-likelihood analysis implemented in the \textit{gtlike} tool.

The model for which we calculated the likelihood is the combination of
point-like and diffuse sources with a  region of interest (RoI) having a
radius of 12$^\circ$  and  centered on the source under consideration. For each point-like
source a power law spectrum  ($F = K E^{-\Gamma}$) derived  from the Fermi Large Area Telescope First Source Catalog (1FGL Catalog;\footnote{http://fermi.gsfc.nasa.gov/ssc/data/access/lat/1yr\_catalog/} Abdo et al.\ 2010b) was adopted. 
Both parameters were allowed to freely vary.
The local model for each RoI includes also sources  falling between 12$^\circ$ and 19$^\circ$
of the target source, which can contribute at low energy due to the broad point-spread function. For these additional
sources, spectral slopes and normalizations were fixed to the values 
provided by the 1FGL catalog.  
As a further check, we repeated the same analysis using a RoI of 10$^{\circ}$. In this case all the sources falling between 
10$^\circ$ and 15$^\circ$ were included and  photon indices of  the point-like  sources fixed to $2$.  
The two different approaches produced completely consistent results.

The background diffuse model used
in the analysis is a combination of the  Galactic emission model
(gll$\_$iem$\_$v02.fit)  and the extragalactic  
and instrumental background (isotropic$\_$iem$\_$v02.txt).\footnote{http://fermi.gsfc.nasa.gov/ssc/data/access/lat/BackgroundModels.html}  The background normalization was allowed to vary freely.
The Galactic emission strongly affects analysis of sources located near the Galactic plane, as the diffuse
background flux is very strong and structured, particularly below 1~GeV.  
In our sample, however, 3C~111 is the only source where the high Galactic background was important in the analysis.

The \textit{gtlike} tool provides the best-fit parameters for each source and 
the significance of each source is given by the test statistic  $TS =$ 2$\Delta$log(likelihood)  between models with and without the source.  
When $TS\le10$, the flux values at $F>100$ MeV
are replaced by $2\sigma$ upper limits, derived by finding  the point 
at which  $2\Delta$log(likelihood)~$=$~4 when increasing the flux from the maximum-likelihood value.

Once the likelihood analysis was performed on the entire 0.1 -- 100 GeV band,  this energy range 
was split into 1, 2  or 3 logarithmically spaced bins per decade, depending on the global flux of the source. 
The flux in each bin was obtained by fitting a power law, while keeping the spectral slope fixed at the value obtained 
by the fit in the entire energy range.As the considered energy bands are small,  it can happen that  a source  has $TS\le10$ in more than one bin in spite of a relatively well constrained spectral shape on the overall 0.1-100 GeV band. 

The departure of the source spectrum from a power law, obtained using the fluxes in the bins, 
was estimated using a $\chi^2$ test following the procedure described in Abdo et al.\ (2010b).

The light curve of each source was generated dividing the total observation period in 15 and 5 
time intervals of 1 month and 3 months duration, respectively,
 and repeating the  likelihood analysis for each interval. The spectral
index of each source was frozen to the best fit over the full interval with the exception of NGC1275,  for which both spectral parameters could also be well constrained in short time intervals. 
A standard $\chi^2$ test was successively  applied to the average flux in each light curve.  We define a source 
as variable if  the probability that its flux is constant is less than $10^{-3}$.
Note that 
all errors reported in the figures or quoted in the text are 1$\sigma$ statistical errors. The estimated systematic errors on the flux,  10\% at 100 MeV, 5\% at 500 MeV and 20\% at 10 GeV,  refer to uncertainties on the effective area of the instrument (see
the 1FGL Catalog; Abdo et al.\ 2010b).


\section{Results}

Table 2  summarizes the results of our analysis.  
For each source, the statistical significance over 15 months of observation 
is listed, in addition to the spectral parameters of the fitting technique
described in the previous section. These include  
the power-law spectral slope, $\Gamma$, the integrated flux, F(E$>$100 MeV), hereafter denoted as
 $F_{100}$ in units of photons cm$^{-2}$ s$^{-1}$, and the derived statistical uncertainties on these parameters.
The K-corrected luminosity L$_{\gamma}$  (erg s$^{-1} $)  between E$_1$=100 MeV and E$_2$= 10 GeV is calculated 
from the relation
$$L_\gamma=1.6 \times 10^{-6} \times 4\pi d_L^2E_1 \frac{(1-\Gamma)}{(2-\Gamma)} \frac{[(\frac{E_2}{E_1})^{2-\Gamma}-1]}{[(\frac{E100}{E_1})^{1-\Gamma}-1]} \times F_{100} $$
where  $d_L$ is the luminosity distance in cm, E100=100 GeV.
Our results are consistent with those reported in the 1LAC catalog. Although not particularly bright, 
with average fluxes of  $F_{100}\sim 6 \times 10^{-8}$, all sample sources  have $TS>30$, 
implying  $\gtrsim 5 \sigma$ detection. Note that the significance (TS)  is dominated by 1-10 GeV photons (see also figure 18 -  $dTS/dlogE$  versus energy in 1FGL  paper) 
while the flux uncertainties by low energy events (characterized by a broad PSF and high background).   
This explains the presence in Table 2 of sources with high significance but large flux uncertainties.

The SEDs of the MAGNs are shown in Figures  1 and 2.   
Seven  of the eleven sources have larger spectral slopes than 2.3,
  so that most photon energies lie between $\approx 100$ ~MeV and 
10~GeV.  
The spectrally softest case is 3C~120, from which only 100 MeV -- 1 GeV emission is detected.  
In spite of its faintness,  the  detection of 3C 120 is however significant at  $\approx 5.6 \sigma$ (Table 2).  
Its analysis  was performed following the  standard procedure but with its position fixed to the optical nuclear coordinates.
Although 3C~120 could be slightly contaminated by the nearby  FSRQ 1FGL J0427.5+0515,  we estimate the effect 
negligible, since the two sources are 1.4$^\circ$ apart and clearly resolved in the count map  (Figure 3).  
Only two FRI radio galaxies, 3C~78 and PKS~0625$-$354, are not detected at
energies $\lesssim 300$ MeV  and show  rather noisy  spectra. These are the weaker sources in the sample with  $F_{100}<10^{-8}$.  
Indeed, for PKS~0625-354 we were forced to restrict the likelihood analysis to the 300~MeV -- 100 GeV band 
in order to constrain the spectral parameters. Due to the low quality of the data,  the SEDs of 3C ~78 and PKS~0625$-$354  are not shown in Figure 1.
3C~111 is another source for which it was necessary to perform a more accurate analysis. As already mentioned, 
this radio galaxy is located at low Galactic latitude and is affected by high background. 
In addition, it was probably in a high state only for a short time during the 
15 months of integration (as briefly discussed later).  
As the source was below the LAT detection threshold for  most of the time,
 a 15-month-integration time necessarily reduces
the source excess counts. Therefore, 
in order to constrain $\Gamma$, all the spectral parameters of the sources within the RoI were initially fixed.
Successively, the  uncertainty in $F_{100}$ was estimated by freezing the spectral slope of 
the target and allowing the normalization and slope of the other point-like sources to vary. 

NGC~1275 is the only source showing a complex spectral shape (see Figure 1), based on the 
larger data set than used in the four month analysis in the original discovery paper, 
where a power-law fit was acceptable (Abdo et al.\ 2009b).
When a power law is applied to the data, the probability that the model is adequate is small 
with $P_{\chi^2} =0.012$, corresponding to $\chi^2=18$ for 7 degrees of freedom (dof).
This radio galaxy  is  also characterized by flux and spectral variability  (see Figure 4).
According to the $\chi^2$ statistics, the model of constant flux ($\chi^2=54$ for 4 dof)  
as well as the model of  constant spectral shape ($\chi^2=28$ for 4 dof) can be ruled out.
A detailed analysis of the spectral evolution of NGC~1275  can be found in Kataoka et al.\  (2010).

Although bright enough to be detected in each temporal bin, no evidence for 
variability was found in our analysis of  M87, Cen A and  NGC~6251. 
However small statistics prevent us to detect factor-of-2 flux changes. 
Incidentally, we note that  NGC~6251  was observed by EGRET   in a brighter flux state (F$_{100} = (74\pm 23)\times 10^{-9}$)
suggesting  a possible $\gamma$-ray variability of this source  on time scales of years (Hartman et al.\ 1999 , Mukherjee et al. 2002). 

The other sources were not significant in each time  interval.  For example, 3C~111 and 3C~120 (Figures 5 and 6),  reached the minimal significance required for detection (TS $>10$) in only one occasion,  even considering a bin integration time of 
3 months. 
In the case of 3C~111  a low duty cycle for $\gamma$-ray emission was also suggested by Hartman et al.\ (2008). 
They noted that this source  only occasionally  became bright ($F_{100} >10^{-7}$) and detectable by EGRET.  
The difference in flux between EGRET and {\it Fermi} detections suggests evidence for long term variability.

\section{Comparison with \emph{Fermi} LAT Blazars}

\subsection{The $\Gamma-L_\gamma$~~Plane}

The spectral slope $\Gamma$  as a function of 0.1 -- 10 GeV $\gamma$-ray luminosity $L_{\gamma}$ 
for all the sources of our sample is plotted in Figure 6, 
along with the corresponding values for 
FSRQs  and BL Lac objects from the 1LAC. 

As can be seen, MAGNs and blazars  occupy different regions of the 
$\Gamma-L_\gamma$ plane. The MAGNs are less luminous on average than 
the $\gamma$-ray blazars.

When a Kolmogorov-Smirnov test is applied, the 
associated probability that blazars and MAGNs are drawn 
from the same population is $P_{\rm KS}<  10^{-3}$.
Although a range of intrinsic source  luminosities can not be excluded, 
this result is in rough accord with expectations from unified scenarios insofar as 
jets that are  not directly pointed towards the observer are expected to 
be fainter because of the smaller Doppler boosting. 

The difference between properties of BL Lac objects and FSRQs on the one hand, and 
FRI radio galaxies and FRII sources on the other, is also evident from Figure 7 where a histogram
of the $\gamma$-ray luminosities of the  FRI radio galaxies and the 
FRII sources  (upper  panel) and  the  BL Lac objects and FSRQs (lower panel) in the 1LAC 
are shown. 
Inspection of Figs. 6-7 shows a well defined separation between FRIs and BL LACs, their putative parent population. On the contrary FRIIs seem to lie 
at best in the outskirts of the FRSQ distribution. 
Although care must be taken to draw conclusions due to the small statistics, 
and about PKS~0943$-$76 because of the uncertainty 
that the Fermi $\gamma$-ray source is correctly associated with it (Section 2), 
a tentative conclusion implied  is that the range of $\gamma$-ray luminosities of 
FRI radio galaxies compared to BL Lacs is larger than that of the FRII galaxies compared to FSRQs.

The significance of this result, if validated with greater statistics as the \emph{Fermi} mission progresses, 
is considered in Section 6.

\subsection{Core Dominance Study of the 3CRR Sample}

In order to better understand the nature of the $\gamma$-ray emitting MAGNs, 
we considered the flux-limited 3CRR sample and analyzed the role of 
core dominance (CD; defined in Section 2), of individual sources that were detected with the 
\emph{Fermi} LAT compared with those that were not detected. We choose the 3CRR sample
because it is well studied and contains the most complete set of data available
for such a study. Because it is restricted to the northern sky, not all the 
MAGNs in our study are considered.
 
Figure 8 shows CD as a function of the total flux density at 178 MHz. 
In this plot, $\gamma$-ray  emitters are identified by triangles inside the filled circles. 
The plot clearly shows  that at a given radio flux, 
radio galaxies and quasars detected at MeV -- GeV energies have the largest CD values. In other words, 
LAT preferentially  selects  the misaligned  AGNs  with smaller angles of inclination.
The MAGNs  radiating at GeV energies do not, however, share the extreme CD values of blazars.  
To demonstrate this, we show the CD values for  two FSRQs,  3C~454.3 and 3C~345, which are 
associated to  {\it Fermi} LAT sources, belong to the 3CRR sample,
and are denoted by the 
blue triangles in black circles in Figure 8 ({\it right panel}). These 
sources occupy  the upper region of the 
CD/$F_{178~MHz}$ plane, much greater than  the values for the misaligned FRII $\gamma$-ray  sources
(which are both, incidentally, the SSRQs).  A similar plot can be obtained by considering the 5 GHz core flux, rather than the total radio emission. 

Considering all the FRI $\gamma$-ray sources listed in Table 1, in particular, 
those shown in  Figure 8 ({\it left panel}), it appears that FRI radio galaxies  with large jet inclination angles and correspondingly small CDs can be observed only if nearby, that is, if  they have large radio
flux densities.  When the distance increases, the source flux becomes weaker, so to have them detectable at GeV energies the CD must increase.

No FRII radio galaxy observed at large angles to the jet axis as reflected by a small CD parameter
has yet been detected with {\it Fermi}.    That we have yet to detect FRIIs with small CD would
simply be a consequence of FRIIs being at larger redshifts than FRIs and thus too weak to be detected.  
Indeed, the median of the 3CRR redshift distribution for FRI and FRII radio galaxies is z= 0.03 and z= 0.56, respectively. The narrow line radio galaxy (NLRG) Cygnus A, 
which is an FRII at
$z=0.056$, exhibits a large core radio flux (F$_c\sim$ 700 mJy at 5 GHz; Hardcastle et al.\  2004), 
of the same order as Cen A. So far, it is not detected with the \emph{Fermi} LAT.
Both objects are seen at large inclination angles, yet both the core and lobes of Cen A are
detected (Abdo et al. 2010c,d).  If proportionality between core radio and $\gamma$-ray flux is assumed 
(see, for example, Giroletti et al.\ 2010),  the origin of the different $\gamma$-ray 
behaviors may reside in the different jet structure.  
While the Cen A jet  could be either decelerating or surrounded by slower external layers, 
Cygnus A might have a collimated relativistic jet with a slower sheath.

\section{Discussion}

Following the Fanaroff-Riley (1974) classification, we have assigned  
the MAGNs into two radio morphological classes
corresponding to  edge-darkened (FRI) and edge-brightened (FRII) objects, with FRII objects 
being more powerful  
(P$_{178~MHz}>10^{25}$ W Hz$^{-1}$ sr$^{-1}$) than FRI radio galaxies. 
In the FRIs, the jets are thought to decelerate and become sub-relativistic on scales of 
 hundreds of pc to kpc, while the jets in FRIIs are at least moderately relativistic and 
supersonic from the core to the hot spots.   
The nuclei of FRIs  are not generally absorbed and are probably powered by inefficient accretion flows 
(e.g. Chiaberge et al.\ 1999, Balmaverde et al.\ 2006).  On the contrary, 
most FRIIs are thought to have an efficient engine and a dusty torus (e.g. Belsole et al.\ 2006).   

From the optical point of view, FRII radio galaxies with bright continuum and broad optical emission lines are classified as 
broad-line radio galaxies (BLRGs). They are classified as narrow line radio galaxies 
(NLRGs)\footnote{With NLRGs we only consider FRII radio galaxies with 
high excitation lines, characterized by an [OIII] equivalent width larger  than 10 \AA~ and/or a
[OII]/[OIII] ratio larger than 1 (Jackson $\&$ Rawlings 1997).} if  their  continuum is weak and only 
narrow  emission lines  are observed.  In the framework of blazar unification, the transition from
NLRGs to BLRGs would represent increasing alignment of the observer along the jet axis.  
A MAGN is defined as a quasar if its integrated optical luminosity is dominated by a point-like source rather than by 
the host galaxy. SSRQs would be high luminosity counterparts of BLRGs.

Assuming isotropic emission in the comoving jet frame with a power law spectrum of index $\alpha$,   
the observed flux density $F(\nu)$  produced through synchrotron and synchrotron self-Compton (SSC) emission
is related to  the rest-frame flux
density  $F^\prime(\nu^\prime )$  through the relation 
$F(\nu)=\delta^{3+\alpha} F^\prime(\nu)$, where $\delta$ is 
the Doppler factor defined by  $\delta=[\gamma(1-\beta \cos\theta)]^{-1}$, 
and $\nu = \delta \nu^\prime/(1+z)$. Here $\beta c$ is the bulk velocity of the emitting plasma, 
$\gamma=(1-\beta^2)^{-1/2}$ is the corresponding Lorentz factor, and $\theta$ is the jet viewing angle (Urry \& Padovani 1995).  A radio-loud AGN  with  $\alpha\approx 1$ and $\gamma \approx 10$  would then have a flux density 
at  $\theta\sim 10^\circ$ smaller than an aligned  blazar by a factor $\approx 250$.  
The Doppler boosting is stronger and the $\gamma$-ray beaming cone narrower compared to 
synchrotron processes if the emission is 
due to Compton scattering of external photons (EC) in the jet. 
In this case the beaming factor of the flux density varies by a factor of $\delta^{4+2\alpha}$ (Dermer 1995), 
or a factor $\delta^{1+\alpha}$ times the synchrotron beaming factor.

Although the number of sources is small, 
the behavior shown in Figure 6 could be explained if the $\gamma$-ray beaming cones 
are narrower in FRII sources, relative to the radio synchrotron beaming
cones, than in FRI sources. This is consistent with the beaming factors just described, provided that
the $\gamma$-ray emission from FRIIs originates from EC processes, 
as would be expected since these sources have prominent broad line regions. 
In this case,  
the stronger reduction in the EC flux for FRII radio galaxies viewed slightly away from the jet axis ($\theta \gtrsim 1/\Gamma$)
as compared to the SSC flux for off-axis FRI radio galaxies makes the detection of off-axis FRI galaxies more probable.
This is in accord with statistical models (M\"ucke \& Pohl 2000; Dermer 2007) of radio galaxies and blazars 
that  predicted a larger number of FRI than 
FRII radio galaxies would be detected by \emph{Fermi} due to the different beaming factors. 

However, the situation is undoubtedly more complicated.
First note that given the strong Doppler boosting of the jetted radiations from blazars, 
the EGRET detection  of the radio galaxies Cen A and  NGC~6251 
(Sreekumar et al.  1999 , Mukherjee et al. 2002) is already somewhat surprising.
SEDs of FRI radio galaxies such as NGC 1275 (Abdo et al.\ 2009b) 
and M87 (Abdo et al.\ 2009c) are consistent with an SSC model with $\Gamma$ (and $\delta)\lesssim 3$, 
which are much lower than typical values found in models of BL Lac objects
(e.g., Costamante \& Ghisellini 2002; Finke et al.\ 2008). 
The corresponding off-axis synchrotron and Compton fluxes
imply a slower bulk velocity of the emitting plasma for the viewing 
angles inferred  by VLBI observations; 
otherwise the debeamed radiation would be much weaker than observed 
(Guainazzi et al.\ 2003; Chiaberge et al.\ 2000, 2001, 2003; Foschini et al.\ 2005). 
This has led to the development of complex models for the structure of the jet, 
including decelerating jet flows  (Georganopoulos \& Kazanas  2003) and the 
spine-layer jet model  (Stawarz \& Ostrowski 2002; Ghisellini et al.\ 2005).  
A possible spine-layer morphology of the jet is also supported  by  earlier 
radio-optical observations concerning polarization properties and 
intensity brightness profiles in both FR I and FR II sources (e.g., Owen et al.\ 
1989, Laing 1996, Swain et al.\ 1998, Attridge et al.\ 1999), as well as 
by numerical simulations of relativistic flows (e.g., Aloy et al.\ 1999, Rossi et al. 2008).

In the MAGN sources that do not exhibit strong evidence for
variability (all except NGC 1275), the 
observed $\gamma$-ray emission could be made 
in regions well beyond the pc scale. For example, a slowly varying,
high-energy emission component can be 
formed by Compton-scattered ambient photon fields, including the CMB radiation (B\"{o}ttcher et al. 2008). 
Proton synchrotron radiation from ultra-relativistic protons in the mG fields of knots and hotspots could also be made at kpc scales from the nuclei of radio galaxies (Aharonian 2002). 
Rapid variability does not necessarily exclude emission from sites outside the pc-scale core, as indicated in X-ray studies of M87 (Cheung et al. 2007; Harris et al. 2009) and Pictor A (Marshall et al. 2010).
A detailed statistical study of Fermi AGNs will be required to test blazar unification, rule out simple one-zone models, and determine the location of the emission region.

We note that the possible detection of several FRIs and some BLRGs 
at GeV energies had already been predicted  before the {\it Fermi} launch. 
In particular, see the papers by Stawarz et al.\ (2003, 2006), Ghisellini et al.\ (2005), 
and Grandi \& Palumbo (2007).

\section{Summary and Conclusions}

We have presented  an analysis of 15 months of LAT  for 
11 radio sources listed in the low radio frequency 3CRR, 3CR and  MS4 catalogs.
In addition to BL Lac and FSRQ blazars,  misaligned radio sources  represent a 
new and important class of  GeV emitters. Among the misaligned AGNs 
studied in this paper, Cen A, NGC~6251, 3C~111 are the only radio galaxies 
that were EGRET candidate sources
(Sreekumar et al. 1999;  Mattox, Hartman, \& Reimer, 2001; Mukherjee et al.\ 2002; 
Sowards-Emmerd, Romani \& Michelson 2003, Sguera et al.\ 2005, Hartman et al.\ 2008). 
The other eight objects  represent new discoveries made with the \emph{Fermi} LAT. 
Dedicated papers have been recently published or are in preparation 
on three of these, namely NGC~1275 (Abdo et al.\ 2009b), M87 (Abdo et al.\ 2009c), 
and Cen A (Abdo et al.\ 2010c,d).

The following points outline our results and conclusions:
\begin{itemize}
\item Our sample is dominated by seven nearby ($d \lesssim 250$ Mpc)
 FRI radio galaxies. Four FRII radio sources, including two FRII radio galaxies and two 
SSRQs, are associated with LAT sources at high probability. The most distant
MAGNs are the SSRQs, at $z\approx 0.7$.

\item The misaligned FRII sources, though few in number, 
are somewhat less $\gamma$-ray luminous than their parent 
population of FSRQs, but have comparable average $\gamma$-ray spectral indices. 
The FRI radio galaxies  are significantly  less luminous than their parent population of  BL Lacs, 
in accord with the unification scenario for radio galaxies and blazars.
The SSRQs appear very similar to $\gamma$-ray emitting FSRQs, 
suggesting a more powerful Doppler  boosting when compared with radio galaxies.

\item A simple power law is a good representation of the 15-month data except for NGC~1275, the brightest source in the sample, which requires a spectral softening above $\approx 3$ GeV. 
NGC~1275 is also the only MAGN for which variability on time scales of months is measured.
Comparison between {\it Fermi} and EGRET fluxes suggests variability on time scale of years for NGC~6251 and 3C~111, in addition to NGC~1275.

\item  Core dominance of the 3CRR sample, which includes 5 MAGNs, indicates that {\it Fermi}  preferentially  detects radio sources intermediate between blazars and 
radio galaxies with large jet inclinations to the line of sight. Only the very nearby radio galaxies M87 and Cen A have small core dominance and large jet angles, suggesting that their detection is in large part
a consequence of their proximity. MAGNs at larger distances have larger values of CD. 

\item The small number of FRIIs with LAT associations could be due to the fewer nearby FRII than 
FRI sources, and to different beaming factors of the emission in the jets of FRII and FRI radio galaxies.

\end{itemize}

As the \emph{Fermi} mission continues, more detections of radio galaxies and misaligned AGNs can be expected. 
Joint statistical analysis of \emph{Fermi} AGNs will test the unification hypothesis of radio 
galaxies and blazars, models for jet structure and $\gamma$-ray beaming, and the contribution of MAGNs to 
the extragalactic $\gamma$-ray background radiation.  Such studies could help explain
the reason for the difference between radio-loud and radio-quiet AGNs.

\section{Acknowledgments}
\acknowledgments
The \textit{Fermi} LAT Collaboration acknowledges generous ongoing support
from a number of agencies and institutes that have supported both the
development and the operation of the LAT as well as scientific data analysis.
These include the National Aeronautics and Space Administration and the
Department of Energy in the United States, the Commissariat \`a l'Energie Atomique
and the Centre National de la Recherche Scientifique / Institut National de Physique
Nucl\'eaire et de Physique des Particules in France, the Agenzia Spaziale Italiana
and the Istituto Nazionale di Fisica Nucleare in Italy, the Ministry of Education,
Culture, Sports, Science and Technology (MEXT), High Energy Accelerator Research
Organization (KEK) and Japan Aerospace Exploration Agency (JAXA) in Japan, and
the K.~A.~Wallenberg Foundation, the Swedish Research Council and the
Swedish National Space Board in Sweden.

Additional support for science analysis during the operations phase is gratefully
acknowledged from the Istituto Nazionale di Astrofisica in Italy and the Centre National d'\'Etudes Spatiales in France.

This research has made use of the NASA/IPAC Extragalactic Database (NED) which is operated by the Jet Propulsion Laboratory, California Institute of Technology, under contract with the National Aeronautics and Space Administration.

{}

 \begin{table}
\caption{The Sample}             
\label{table:1}                                                                             
\scriptsize                                                                          
\begin{tabular}{ll l l l c  l l ll }                                                           
\hline\hline                                                                                 
Object  & 1FGL Name & RA         & Dec        &   Redshift             &\multicolumn{2}{c}{Class} & Log (CD) & ref & Cat.\\
              & &   (J2000) & (J2000)   &                               & Radio&Optical                   & at 5 GHz   &      &          \\

\hline
3C~78/NGC~1218  & 1FGLJ0308.3+0403   &03~08~26.2&+04~06~39&	0.029&FRI   &G          &  $-$0.45 &  1               & 3CR  \\
3C~84/NGC~1275   & 1FGLJ0319.7+4130  &03~19~48.1&+41~30~42&	0.018&FRI  & G          &  $-$0.19      &  2$^a$ & 3CR \\
3C~111	                & 1FGLJ0419.0+3811 &04~18~21.3&+38~01~36&	0.049& FRII&BLRG   & $-$0.3         &  3               & 3CRR	\\
3C~120                      &                                         &04~33~11.1&+05~21~16&	0.033 &FRI & BLRG	 & $-$0.15	   & 1              & 3CR\\
PKS~0625-354$$	       &1FGLJ0627.3$-$3530   &06~27~06.7&$-$35~29~15&	0.055&FRI$^b$&G             &  $-$0.42    & 1               & MS4 \\
3C~207     		       &1FGLJ0840.8+1310  & 08~40~47.6&+13~12~24&	0.681&	FRII&SSRQ	 & $-$0.35        & 2                & 3CRR \\    
PKS~0943-76	       &1FGLJ0940.2$-$7605   & 09~43~23.9&$-$ 76~20~11&	0.27&	FRII&	G       & $<-0.56$  & 4                  & MS4\\
M87/3C~274	                         & 1FGLJ1230.8+1223  & 12~30~49.4&	+12~23~28&	0.004&	FRI& G       &$-$1.32         & 2                & 3CRR	\\
Cen A                       & 1FGLJ1325.6-4300   & 13~25~27.6 &$-$ 43~01~09                    &	0.0009$^c$&	FRI &G       &  $-$0.95                 &  1              &  MS4\\
NGC~6251          	&1FGLJ1635.4+8228& 16~32~32 .0&	+82~32~16&	0.024&	FRI&G                 & $-$0.47        & 2               &	3CRR\\
3C~380	&1FGLJ1829.8+4845& 18~29~31.8&	+48~44~46&	0.692&	FRII/CSS  &SSRQ             & $-$0.02        & 2               &  3CRR\\

 \hline                                   
\multicolumn{10}{l}{\tiny {(1) Morganti et al.\ 1993; (2) 3CRR database; (3) Linfield $\&$ Perley 1984; (4) Burgess \& Hunstead (2006a,b)}}\\
\multicolumn{10}{l}{\tiny{$^a$ --  More recent 5~GHz core flux taken from Taylor et al.\ 2006. The CD value is uncertain because of the radio core variability. } }\\
\multicolumn{10}{l}{\tiny{$^b$ --  This source shows some BL Lac characteristics in the optical band (see Wills et al.  2004) } }\\
\multicolumn{10}{l}{\tiny{$^c$ --  The Cen A  distance is assumed to be 3.8 Mpc (Harris et al.\ 2009).}}\\

\end{tabular}
\end{table}

%
\begin{table}
\caption{Results of the {\it Fermi} LAT  analysis}             
\label{table:2}                                                                             
\footnotesize                                                                          
\begin{tabular}{l l l l l c }                                                           
\hline\hline                                                                                 
Object  & TS        & $\Gamma$   &Flux$^a$   & $Log~ Lum^b$ \\
             &                              &                                                           & ($>100$~MeV)                             & (0.1-10 GeV)\\
\hline

3C~78/NGC~1218	                             & 35  	           &  1.95$\pm$0.14   &       4.7  $\pm$1.8           & 	42.84 \\
3C~84/NGC~1275	          & 4802	           &  2.13$\pm $ 0.02 &       222   $\pm$8	        &	44.00 \\
3C~111	                             & 34   	           &  2.54 $\pm 0.19$                 &      40 $\pm8^c$       &        44.00 \\
3C~120	                             & 32	                    &  2.71$\pm $0.35       &      29    $\pm$17       &       43.43 \\
PKS~0625-354$^d$                      & 97	                    & 2.06$\pm0.16$  &      4.8       $\pm1.1$	        &      43.7\\
3C~207 	                             & 79	                    & 2.42 $\pm $0.10      &      24  $\pm$4	         &      46.44 \\
PKS~0943-76                        & 65	                    & 2.83 $\pm $ 0.16  &      55     $\pm$12	          &     45.71\\
M87/3C~274                                       & 194	          & 2.21 $\pm $ 0.14   &      24  $\pm$ 6  	 &     41.67\\
Cen A	                            & 1010	          & 2.75$\pm$ 0.04     &     214    $\pm$12 	          &     41.13\\
NGC~6251                            & 143	          & 2.52  $\pm $0.12	&     36       $\pm$8 	          &     43.30\\ 
3C~380                                  & 95	                   & 2.51    $\pm $0.30      &    31       $\pm$18	          &     46.57\\

 \hline                                   
 \multicolumn{5}{l}{\tiny{$^a$ - $\times ~10^{-9} $ ~Photon~cm$^{-2}$~ $s^{-1}.$}}\\
\multicolumn{5}{l}{\tiny{$^b$ -  erg~s$^{-1}$}}\\
\multicolumn{5}{l}{\tiny{$^c$ -  Flux was estimated keeping the spectral slope fixed.}}\\
\multicolumn{5}{l}{\tiny{$^d$ -  Likelihood analysis was limited to the 300 ~MeV -- 100 GeV range.}}\\ 
\multicolumn{5}{l}{\tiny{Flux ($>300$~MeV); luminosity extrapolated down to 100 MeV}}\\
\end{tabular}
\end{table}
\noindent
\begin{figure}[htbp]
\begin{center}
\plottwo{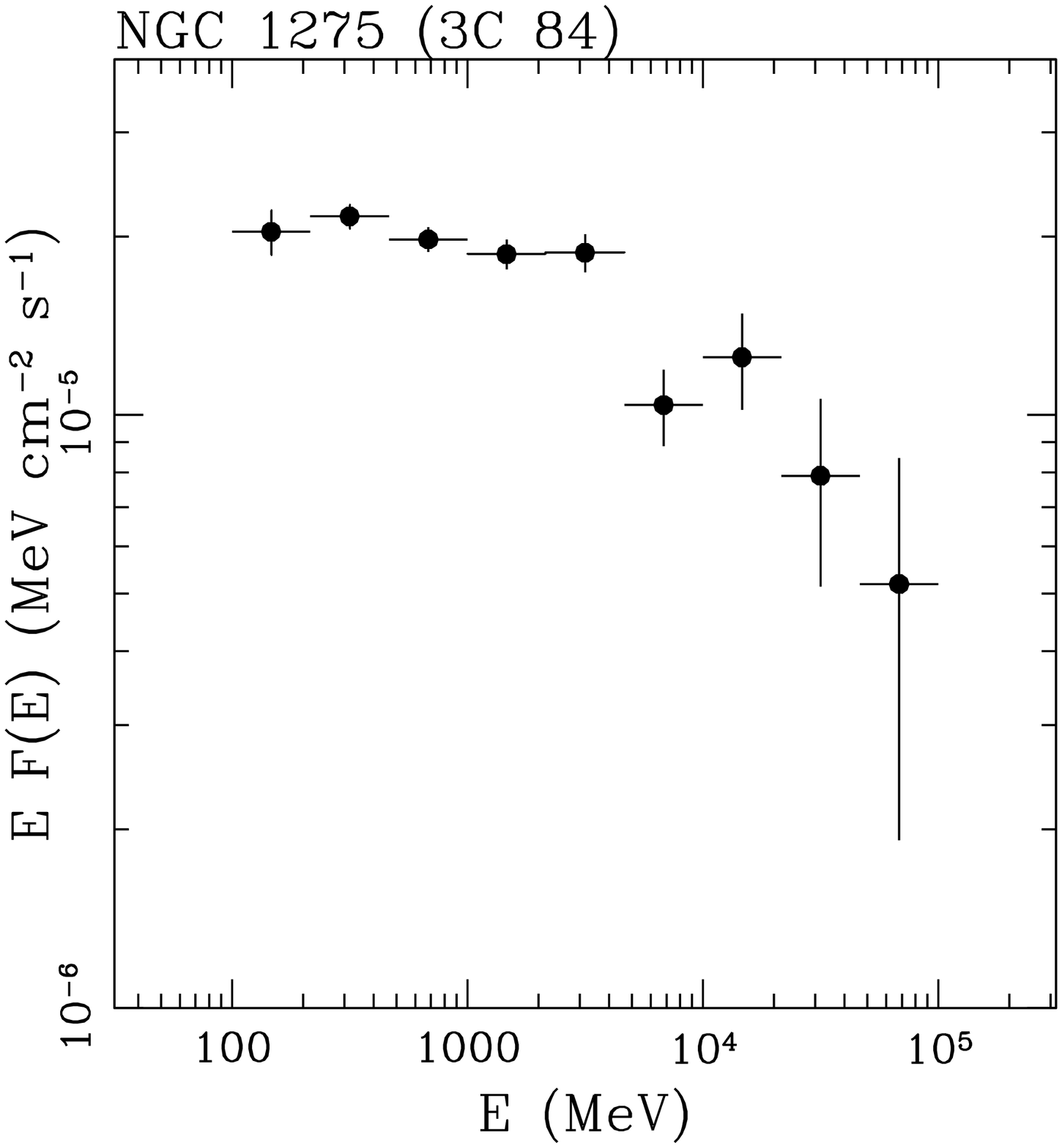}{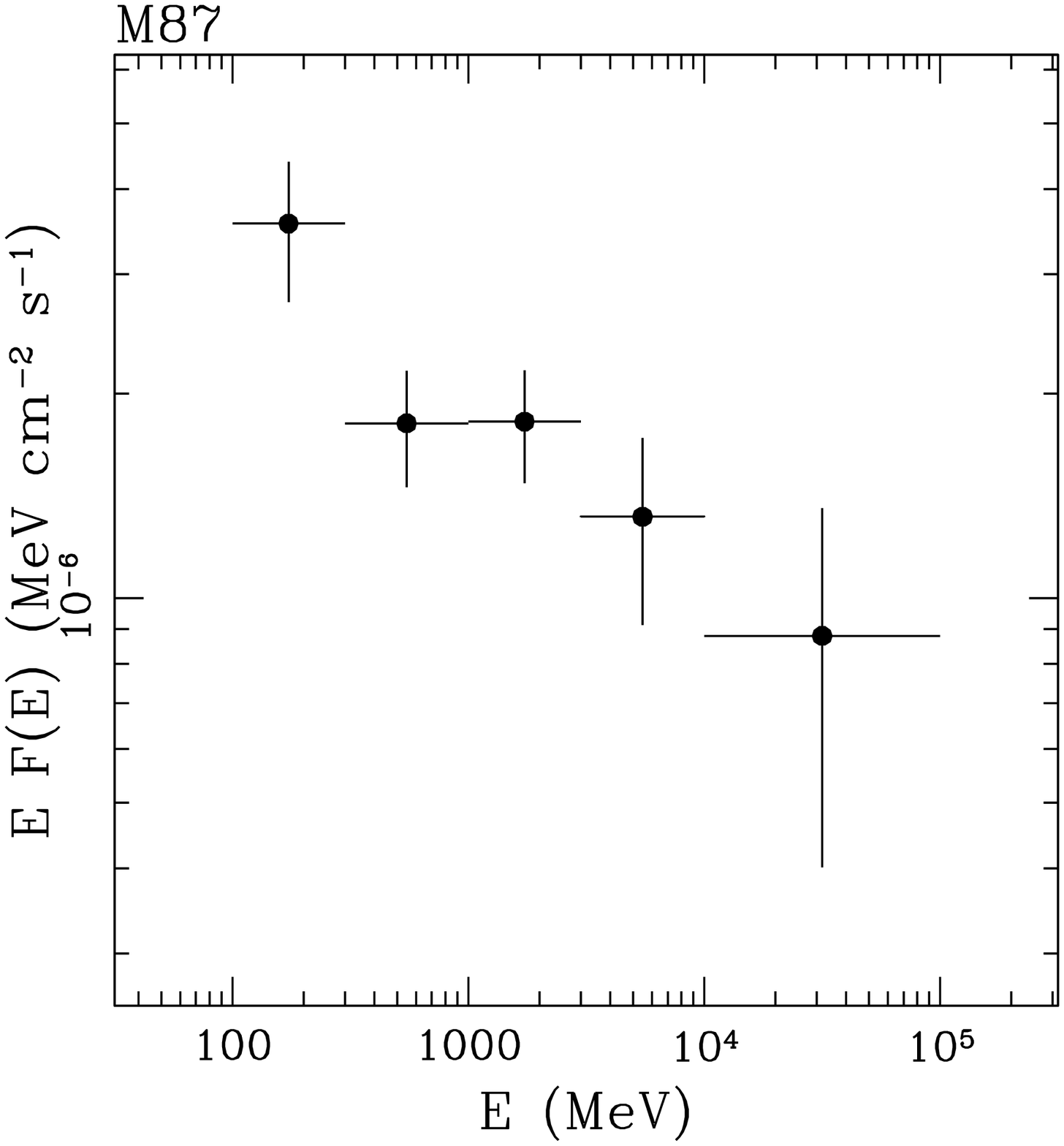}
\plottwo{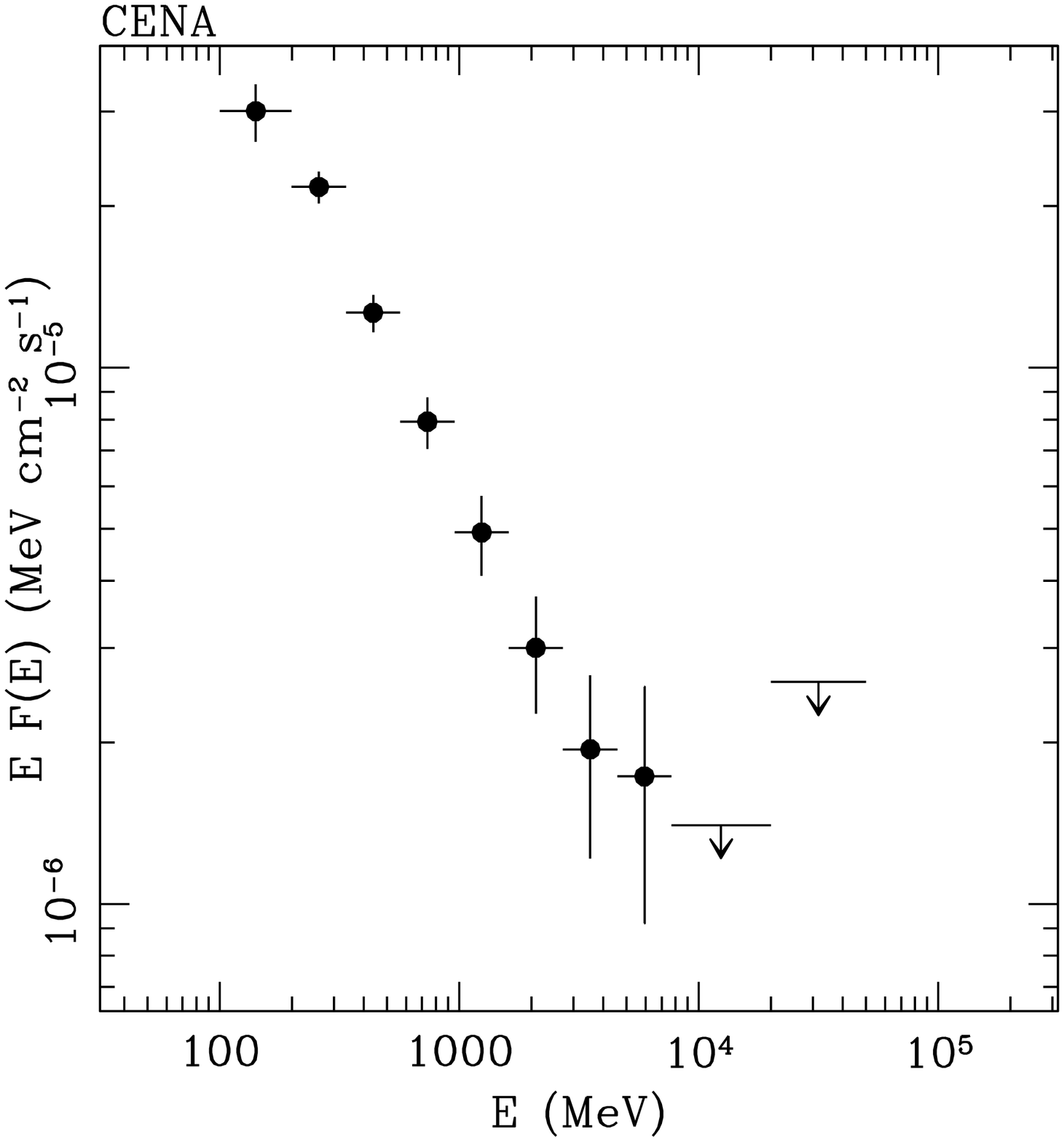}{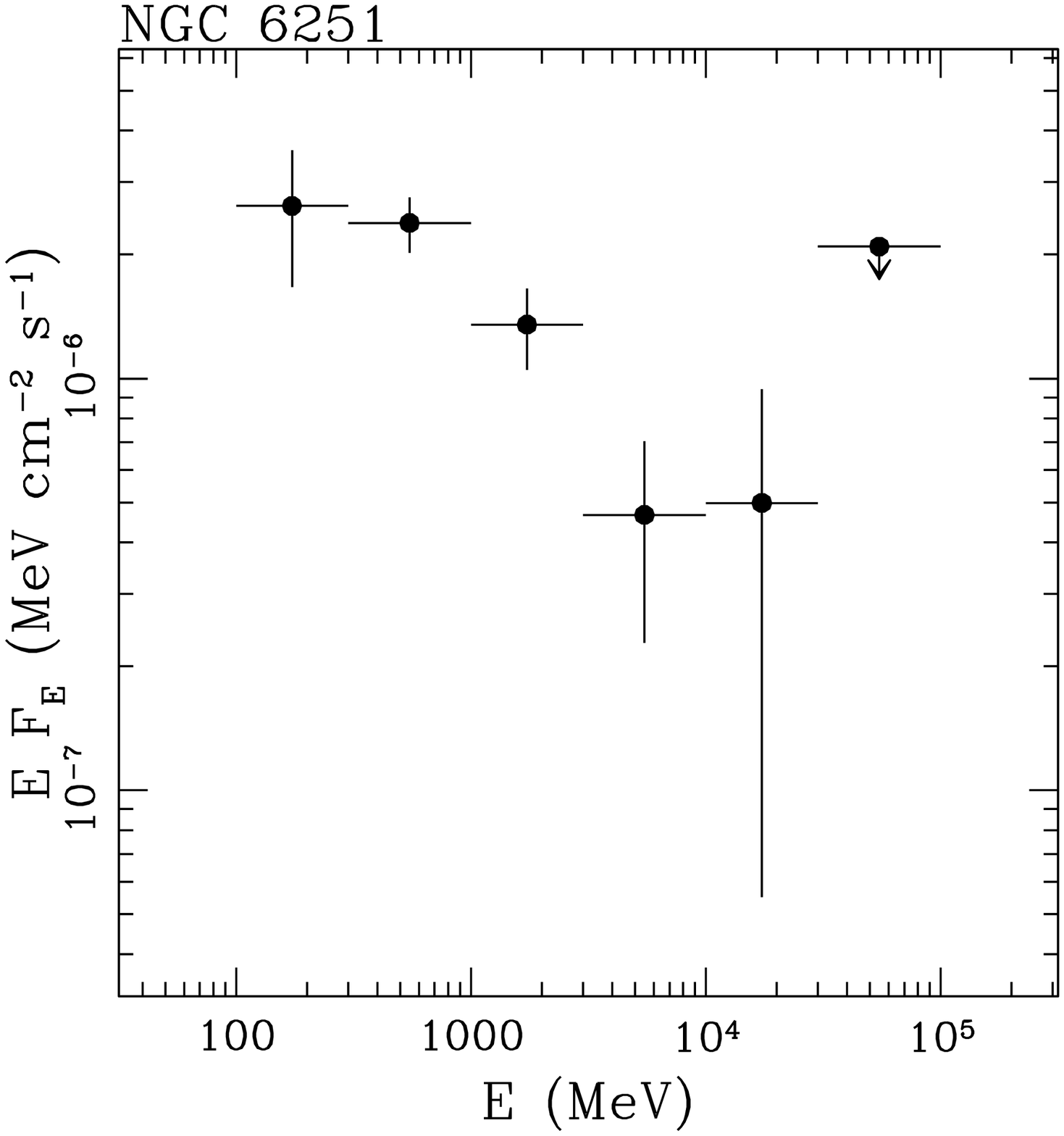}
\caption{{Spectral energy distributions (SEDs) of the FRI radio galaxies NGC~1275, M87, Cen A, NGC~6251. }}
\label{fig1}
\end{center}
\end{figure}

\begin{figure}[htbp]
\begin{center}
\plottwo{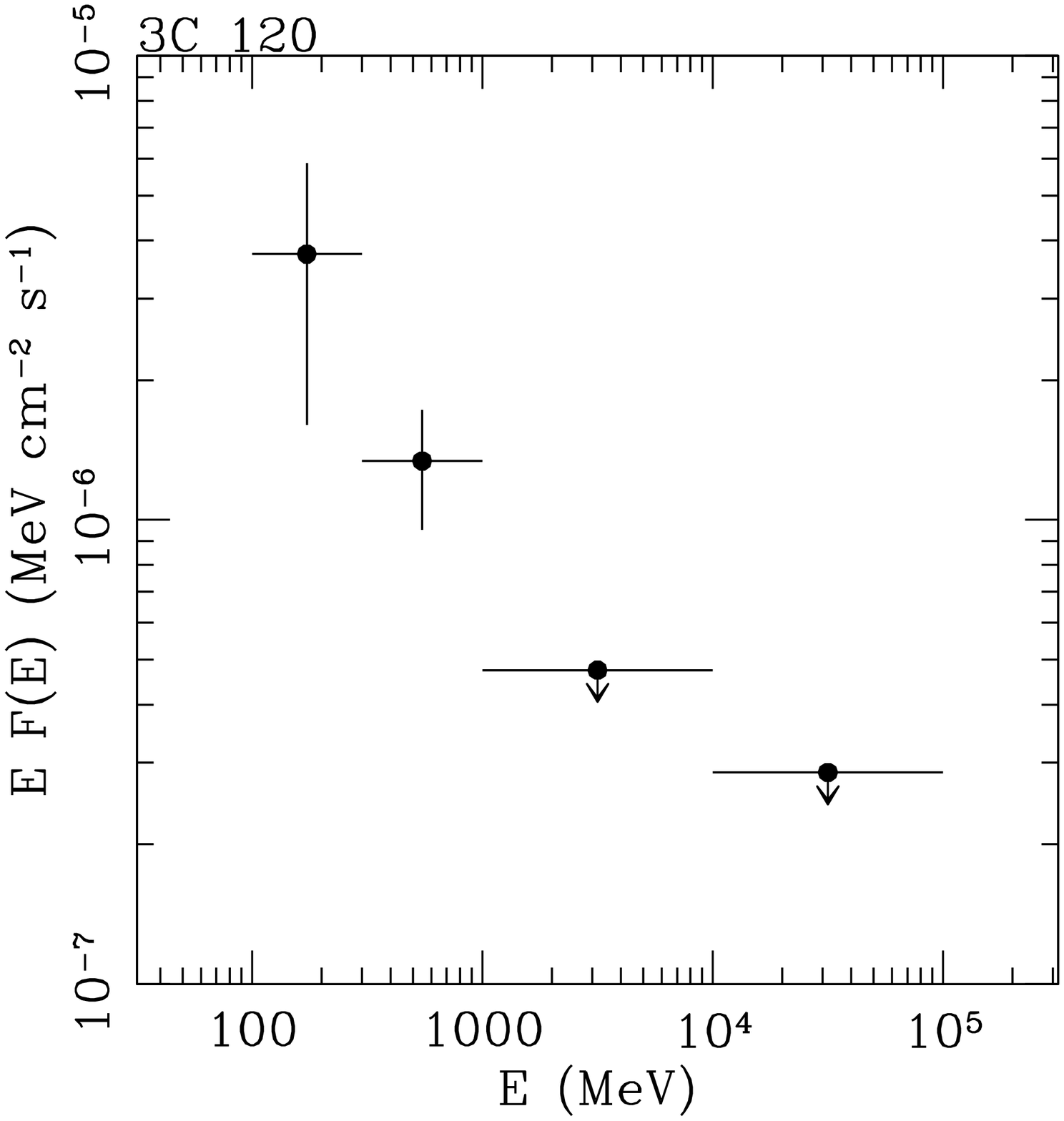}{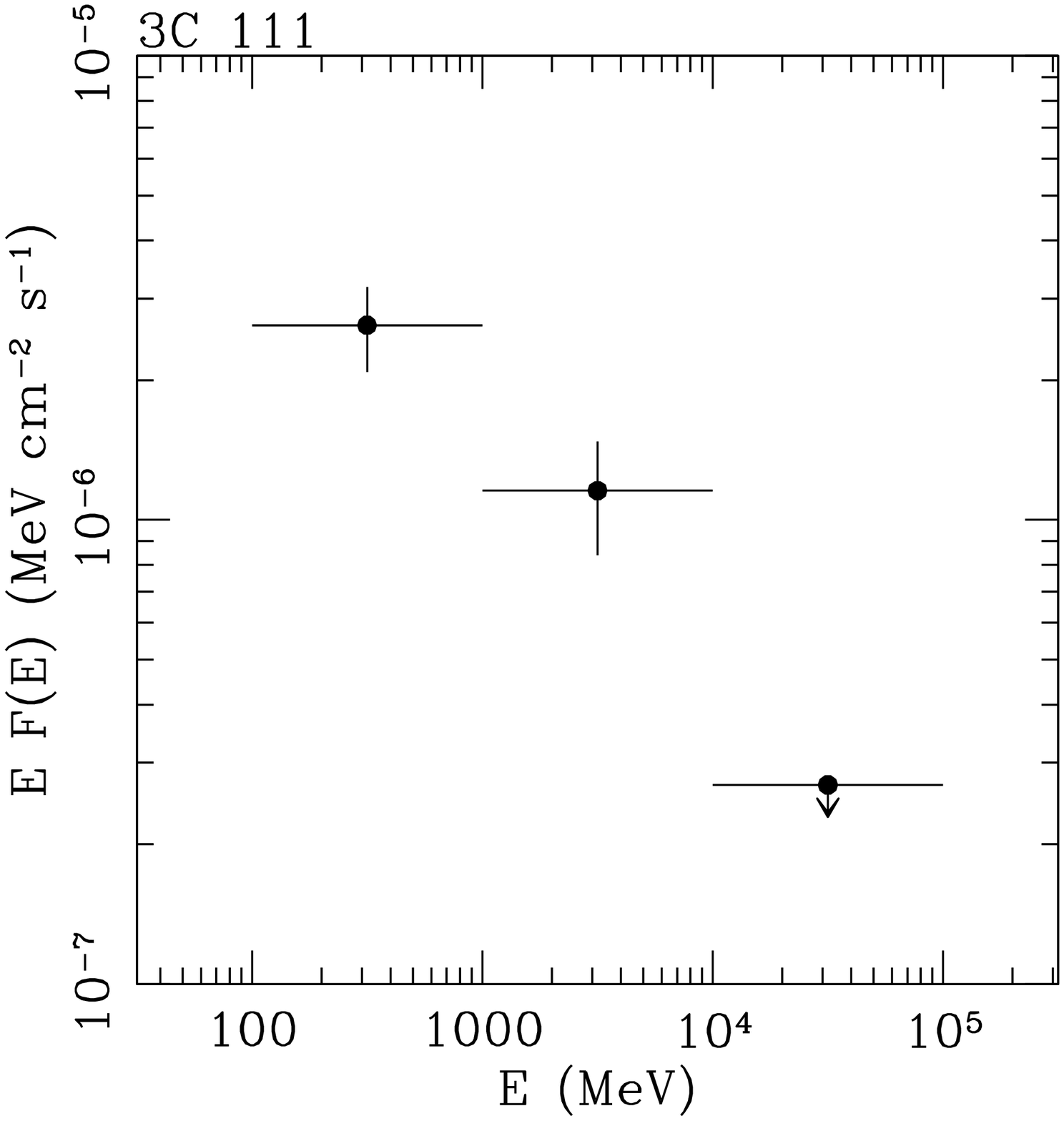}
\plottwo{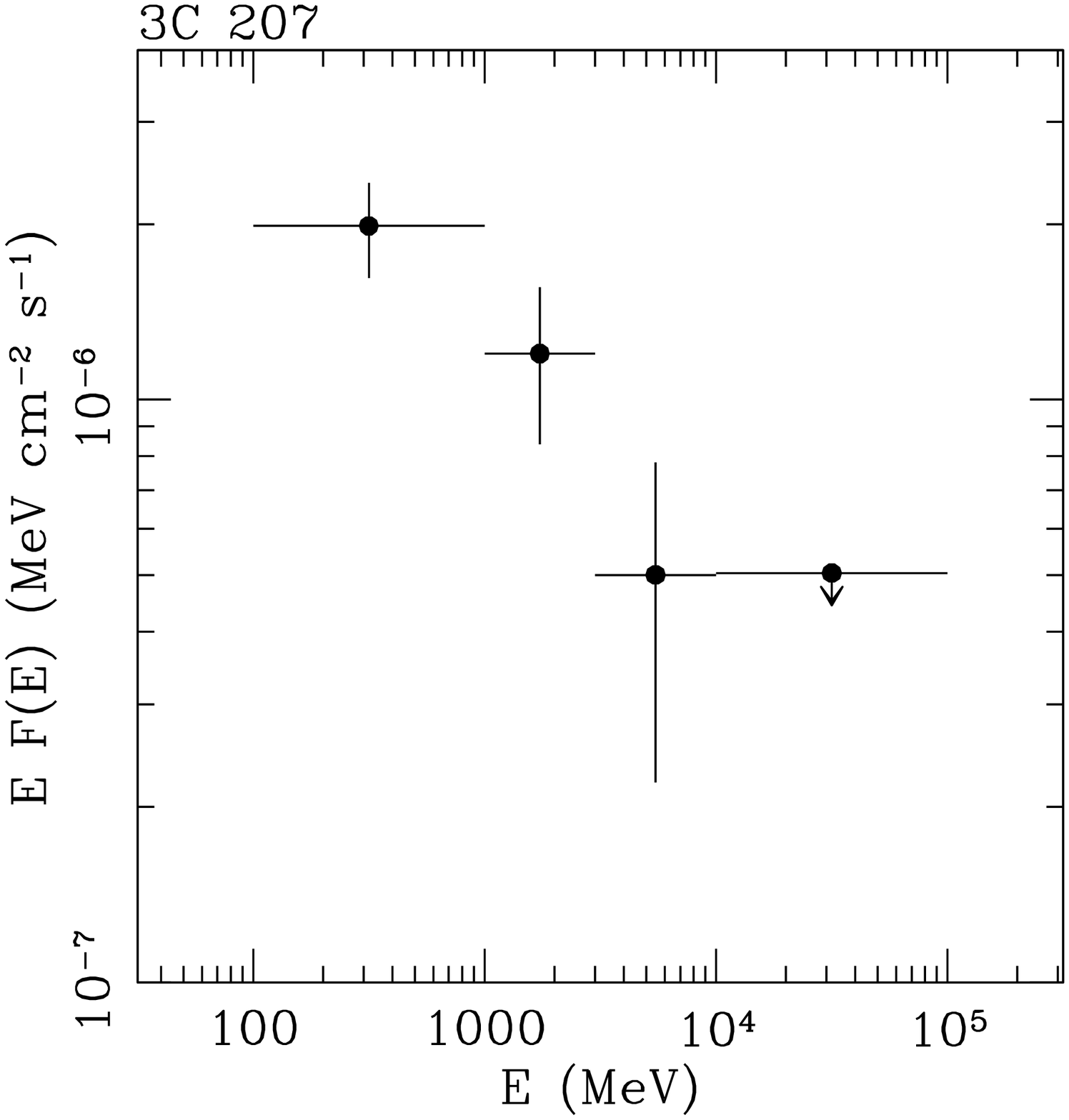}{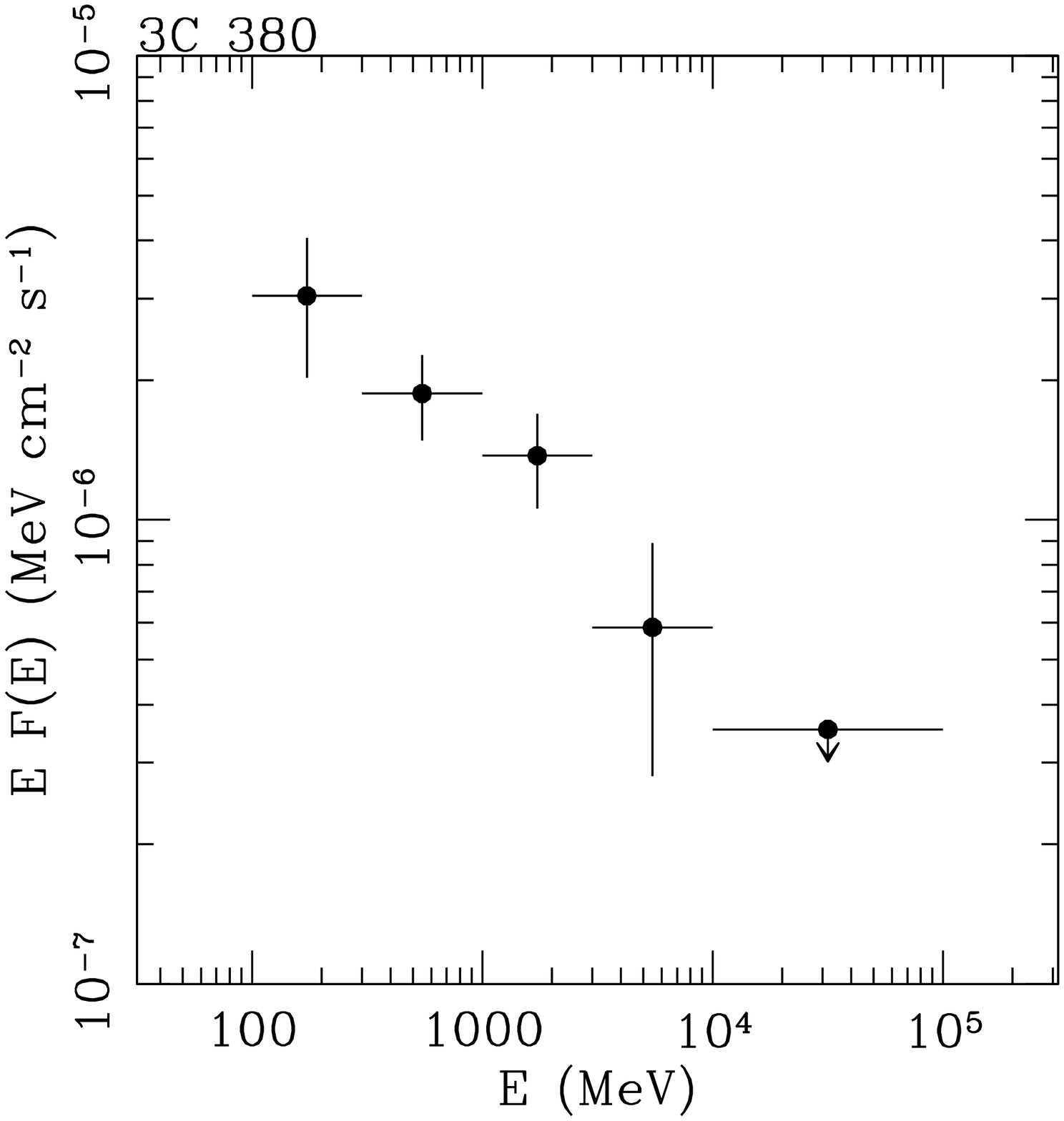}
\caption{{SEDs of the two BLRGs 3C~120 and 3C~111  ({\it upper  panel})  and of the  two steep spectrum radio quasars 3C~207 and 3C~380 ({\it lower panel}). }}
\label{fig2}
\end{center}
\end{figure}

\begin{figure}[htbp]
\begin{center}
\epsscale{0.8}
\plotone{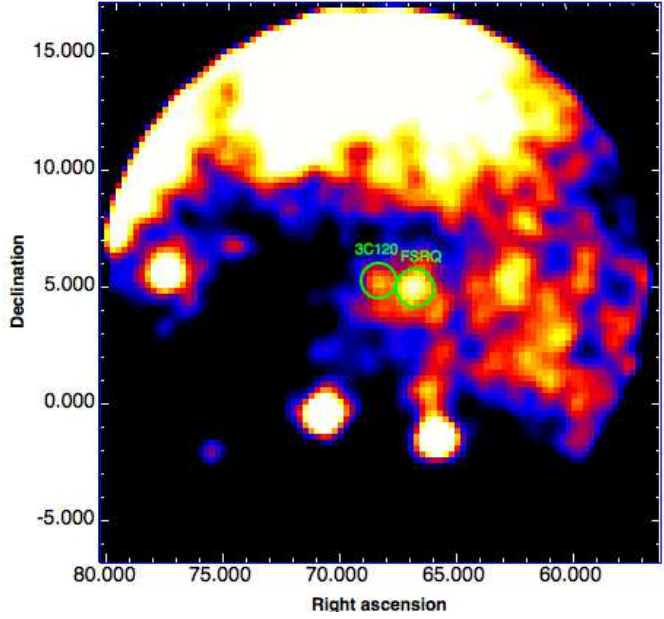}
\caption{{3C~120 count sky map between 100 MeV  and 100 GeV.  }}
\label{fig3}
\end{center}
\end{figure}

\begin{figure}[htbp]
\epsscale{0.5}
\begin{center}
\plotone{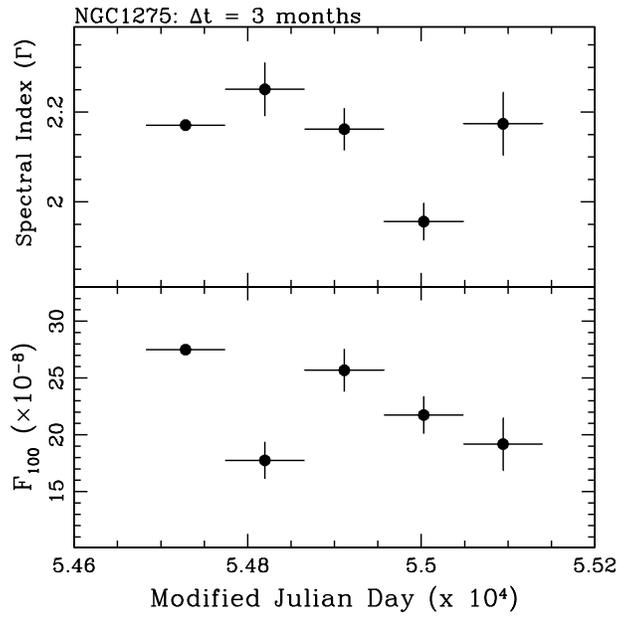}
\caption{{Flux and  spectral slope variations of NGC~1275.  Each bin corresponds to 3 months of observations in the 100 MeV-100 GeV band.}}
\label{fig4}
\end{center}
\end{figure}

\begin{figure}[htbp]
\begin{center}
\epsscale{0.5}
\plotone{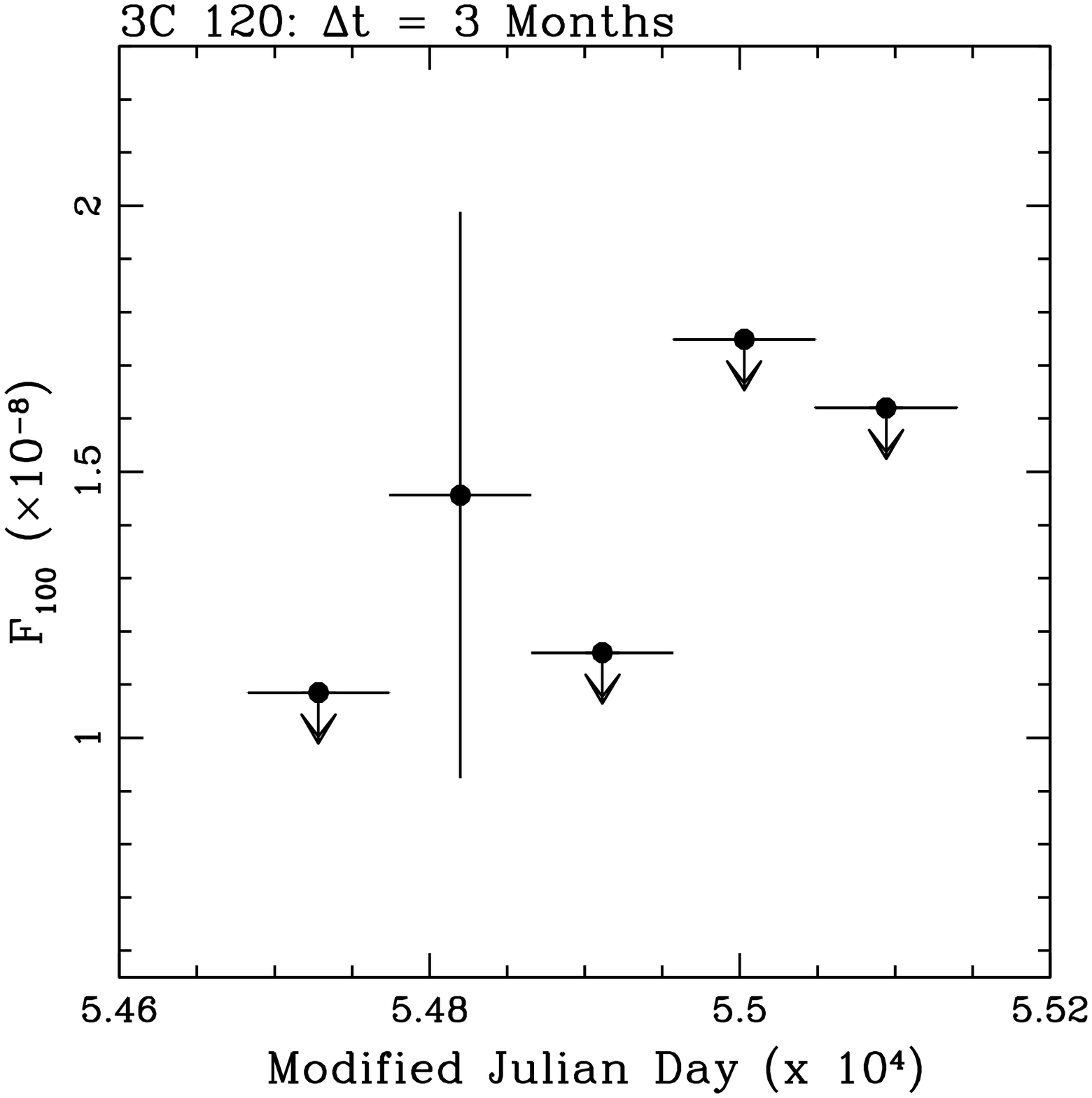}
\plotone{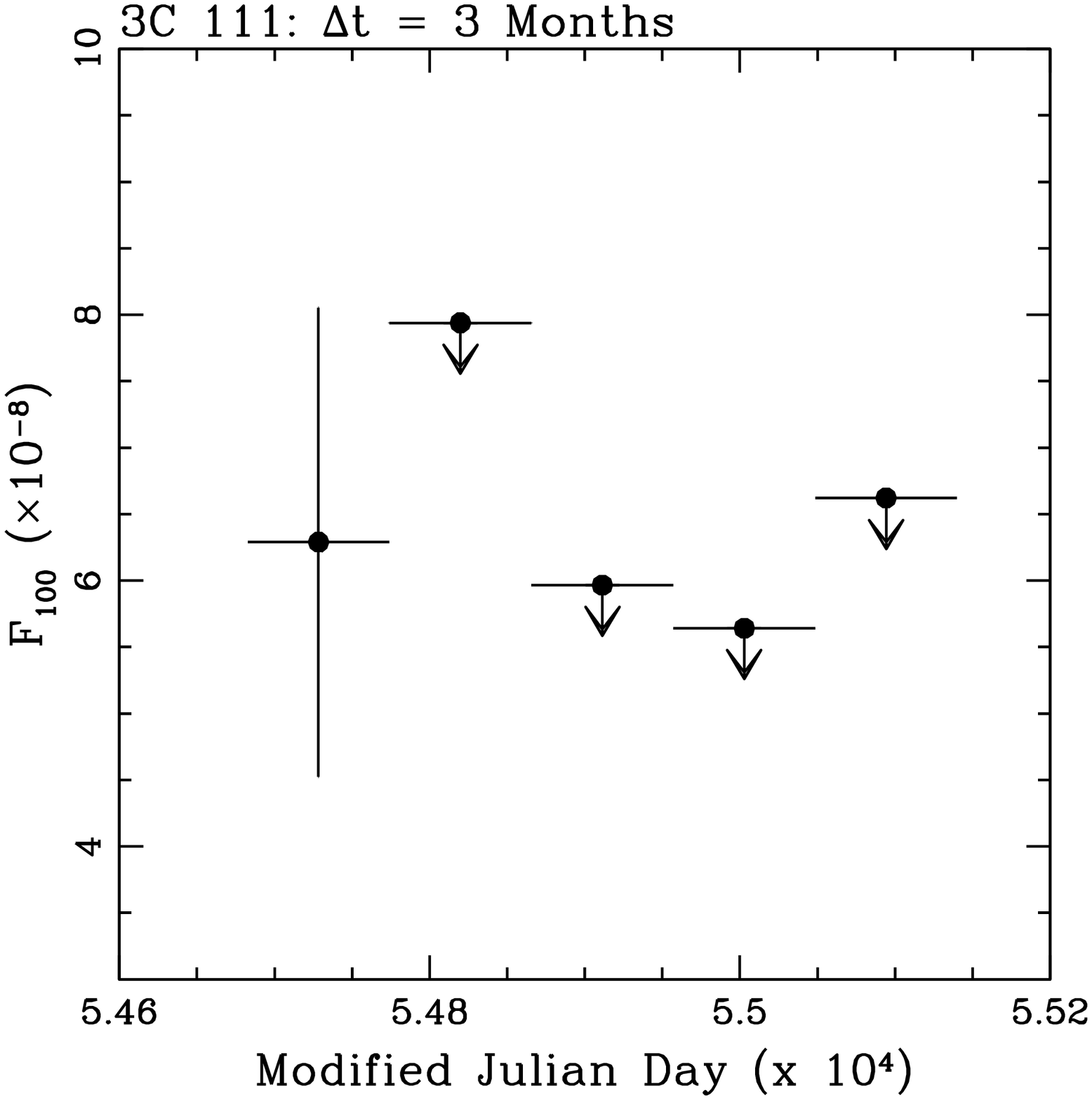}
\caption{{Light curves of the broad line radio galaxies 3C~120 and 3C~111
    between 100 MeV and 100 GeV. Each bin covers three months of observations. }}
\label{fig5}
\end{center}
\end{figure}

\begin{figure}[htbp]
\begin{center}
\epsscale{.80}
\plotone{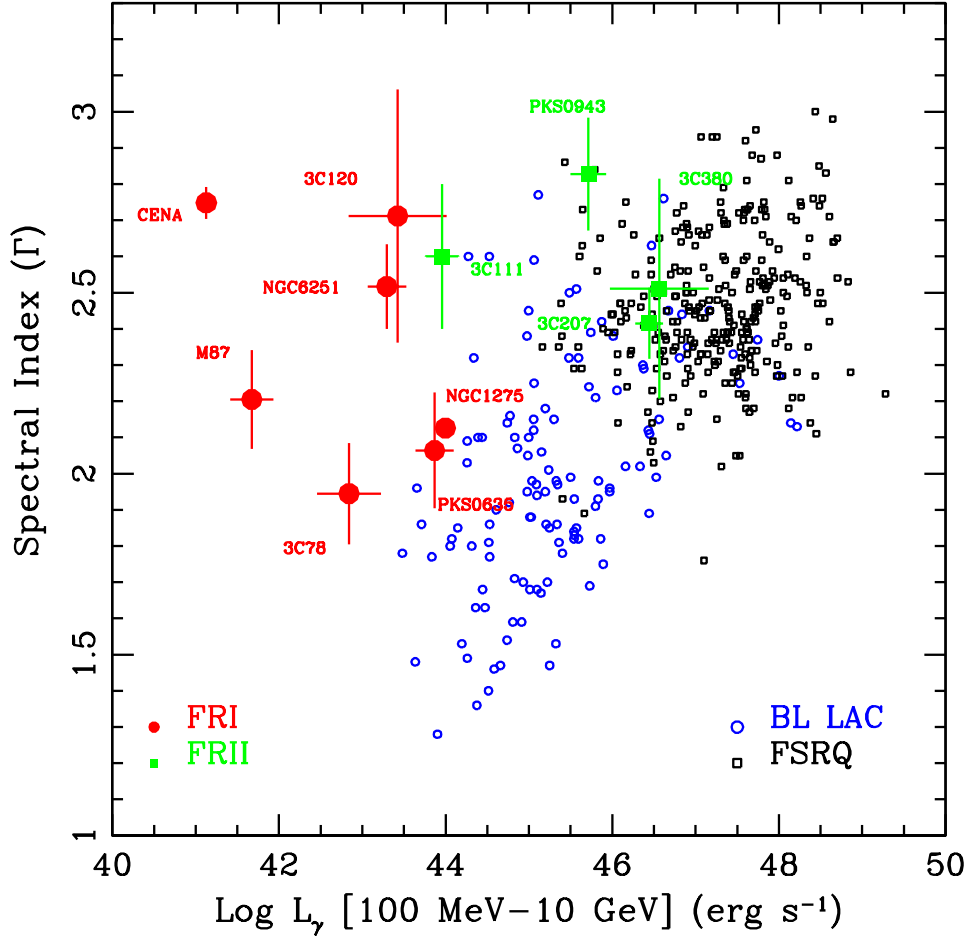}

\caption{ {The spectral slopes of FRI radio galaxies (red circles), FRII radio sources (green squares),  BL Lacs (open blue circles) and FSRQ (open black squares) are plotted as a function of the $\gamma$-ray luminosity (100 MeV-10 GeV).  Local radio galaxies ($z<0.1$) and blazars occupy different regions of the plot, with misaligned AGNs generally characterized by lower luminosity.  On the contrary, the two more distant steep spectrum radio quasars ($z>0.6$) fall within the range of $\gamma$-ray luminosities of FSRQs. }}
\label{fig6}
\end{center}
\end{figure}  

\begin{figure}[htbp]
\begin{center}
\epsscale{0.80}
\plotone{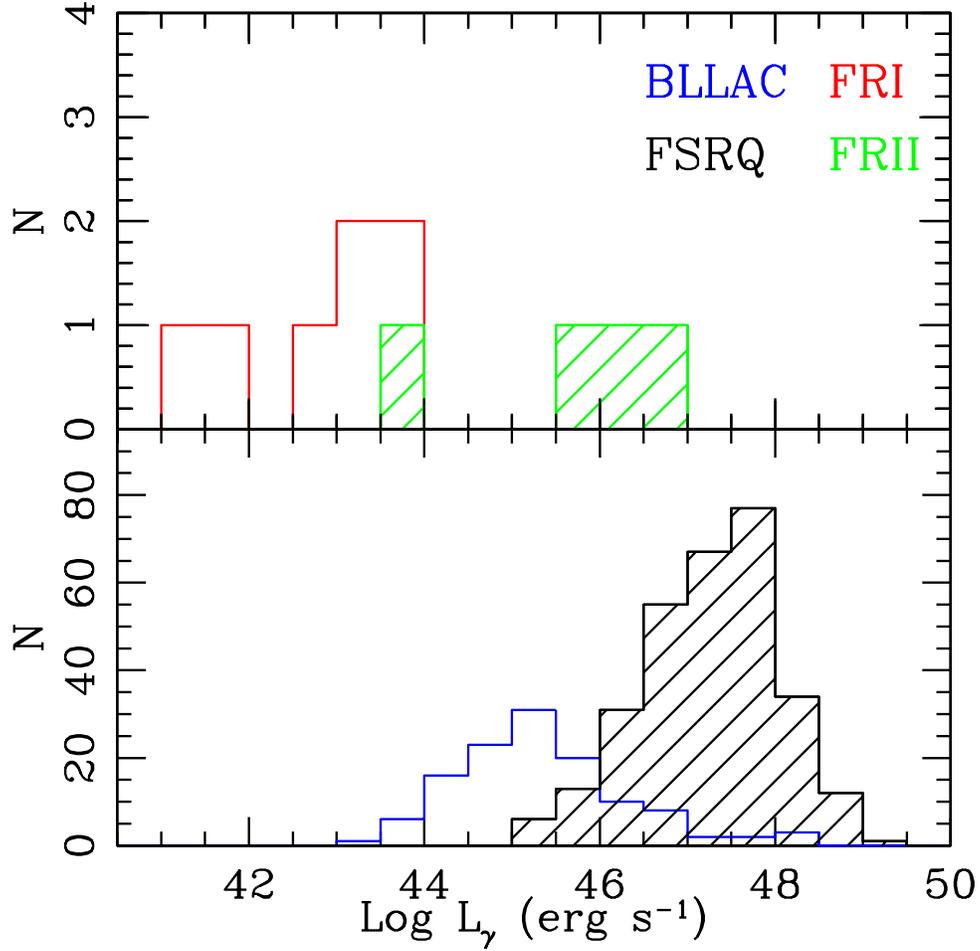}
\caption{{Histogram showing the luminosity distribution of misaligned AGN  ({\it upper panel : FRIs - red continuum line, FRIIs - green dashed line})
and blazars ({\it lower panel: BL Lacs - blue continuum line; FSRQs - black dashed line}).  FRI radio galaxies  are  significantly less luminous than BL Lac objects . The Broad Line radio Galaxy 3C~111 is the only FRII  outside the luminosity range covered by the FSRQ }}
\label{fig7}
\end{center}
\end{figure}

\begin{figure}[htbp]
\begin{center}
\epsscale{0.80}
\plotone{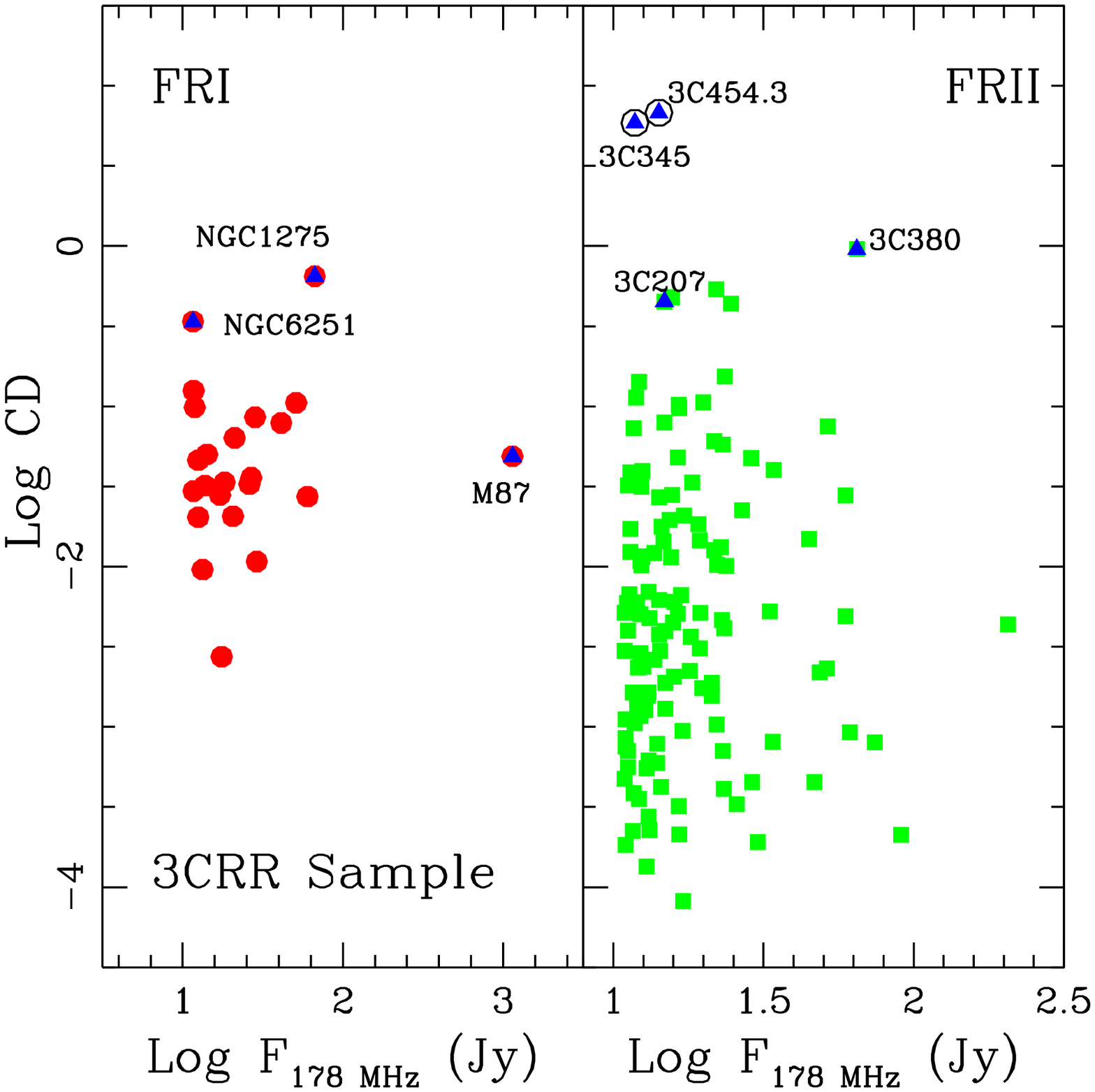}
\caption{{Core dominance (CD)  versus  total flux at 178 MHz of all the sources of the  3CRR sample (FRI - red circles; FRII - green squares) with a measured radio core.  
The MAGN detected by {\it Fermi} (blue tringles in circles/squares)  are characterized by large core dominances. The two FSRQs (blue triangles in empty black circles) belonging to the 3CRR and associated to LAT sources have much larger CD values
than the misaligned FRII sources.}}
\label{fig8}
\end{center}
\end{figure}


\begin{thebibliography}{}

\bibitem{} Abdo, A.A et al. 2009a,  ApJ, 700, 597 (LBAS)
\bibitem{} Abdo, A.A et al.  2009b,  ApJ, 699, 31 (NGC 1275)
\bibitem{} Abdo, A.A et al.  2009c,  ApJ, 707, 55 (M87)
\bibitem{} Abdo,  A.A et al. 2010a,  ApJ, 715, 429 (1LAC)
\bibitem{}  Abdo,  A. A., et al.  2010b,   ApJS, 188, 405 (1FGL)
\bibitem{} Abdo,  A. A., et al.  2010c,  Science, 328, 725  (Cen A Lobes)
\bibitem{} Abdo, A. A., et al. 2010d,  ApJ in press (Cen A core)
\bibitem{} Aharonian, F.~A.  2002, \mnras, 332, 215 
\bibitem{} Aloy, M. A., Ib\'{a}\~{n}ez, J. M., Mart\'{i}, J. M. et al. 1999, ApJ, 523, L125
\bibitem{} Attridge, J. M.,  Roberts, D. H., Wardle, J. F. 1999, ApJL, 518, 87 
\bibitem{} Atwood, W. B., et al.  2009, ApJ, 697, 1071
\bibitem{} Bennett, A. S. 1962, MNRAS, 125, 75
\bibitem{} B{\"o}ttcher, M. 2007,  The Central Engine of Active Galactic Nuclei, 373, 169 
\bibitem{} B{\"o}ttcher, M.,  Dermer, C.~D., \& Finke, J.~D. 2008, \apjl, 679, L9     
\bibitem{} Burgess, A. M. \& Hunstead, R. W. 2006a, AJ, 131, 100
\bibitem{} Burgess, A. M. \& Hunstead, R. W. 2006b, AJ, 131, 114
\bibitem{} Chiaberge, M. et al.\ 2000, A\&A, 358, 104
\bibitem{} Chiaberge, M., Capetti, A., Celotti, A. 2001, MNRAS, 324, 33
\bibitem{} Chiaberge,  M., Gilli, R., Capetti A. \& Macchetto F. D. 2003, ApJ, 597, 166
\bibitem{} Cheung, C.~C., Harris, D.~E., \& Stawarz, {\L}.\ 2007, \apjl, 663, L65 
\bibitem{} Costamante, L., \& Ghisellini, G.\ 2002, \aap, 384, 56 
\bibitem{} Dermer,  C. D. 1995, ApJ, 446, L63
\bibitem{} Dermer, C.~D.\ 2007, \apj, 659, 958 
\bibitem{} Fan, J, H. \& Zhang, J. S.,  2003, A\&A, 407, 899 
\bibitem{} Fanaroff, B. L.,\& Riley, J. M. 1974, MNRAS, 167, 31
\bibitem{} Finke, J.~D., Dermer, C.~D., \& B\"ottcher, M.\ 2008, \apj, 686, 181
\bibitem{} Foschini L., Chiaberge M., Grandi P. et al.\ 2005, A\&A, 433, 515
\bibitem{} Georganopoulos, M. \& Kazanas, D. 2003, ApJ, 589, L5
\bibitem{} Ghisellini, G., Tavecchio, F., Chiaberge, M. 2005, A\&A, 432, 401
\bibitem{} Giroletti, M. Reimer, A.,  Fuhrmann, L., et al.\ 2009 Fermi Symposium, eConf Proceedings C091122 (arXiv:1001.5123v1)
\bibitem{}ÊGrandi, P. \& Palumbo, G.G. C., 2007, ApJ 659, 235
\bibitem{} Guainazzi, M., Grandi P., Comastri A. \& Matt G. 2003, A\&A, 410, 131 
\bibitem{} Hardcastle M. J. Harris, D. E., Worrall, D. M. et al.\ 2004, ApJ, 612, 729
\bibitem{} Harris, D.~E., Cheung, C.~C., Stawarz, {\L}., Biretta, J.~A., \& Perlman, E.~S.\ 2009, \apj, 699, 305 
\bibitem{}ÊHartman, R.C., Bertsch, D. L., Bloom, S. D., et al.\ 1999, ApJS, 123, 79
\bibitem{} Hartman, R.C., Kadler, M \& Tueller, J. 2008, ApJ, 688, 852
\bibitem{}ÊHarris, G. L. H., Rejkuba, M., Harris W. E. 2009, PASA in press ( arXiv:0911.318)
\bibitem{}ÊJackson N., $\&$ Rawlings S. 1997, MNRAS 284, 241
\bibitem{}Ê Kataoka J., Stawarz, L. Cheung, C.C. et al.\ 2010, ApJ, 715, 554
\bibitem{}ÊKellermann, K. I., Pauliny-Toth, I. I. K., \& Williams, P. J. S., 1969, ApJ, 157, 1
\bibitem{}ÊKomatsu, E., et al.\ 2009, ApJS, 180, 330 
\bibitem{}ÊLaing, R. A., Riley, J. M., \& Longair, M. S. 1983, MNRAS, 204, 151 
\bibitem{} Laing, R. A. 1996, {\it Energy transport in radio galaxies and quasars},  Astronomical Society of the Pacific Conference Series, Volume 100, p.241
\bibitem{} Ledlow, M.~J., \& Owen, F.~N.\ 1996, \aj, 112, 9 
\bibitem{}ÊLinfield, R. \& Perley, R. 1984, ApJ, 279, 60
\bibitem{}ÊMarshall, H.~L., et  al.\ 2010, \apjl, 714, L213
\bibitem{}ÊMattox, J. R., Hartman, R. C., Reimer, O. 2001, ApJ S, 135, 155
\bibitem{}ÊMorganti,  R.,  Killeen N. E. B., Tadhunter, C. N.  1993, MNRAS, 263, 1023
\bibitem{} M{\"u}cke, A., \& Pohl, M.\ 2000, \mnras, 312, 177 
\bibitem{}ÊMukherjee, R., Halpern, J., Mirabal,  N. \& Gotthelf E. V. 2002, ApJ, 574, 693
\bibitem{}ÊNeronov, A., Semikoz, D. and Vovk, Ie 2010, A\&A  in press (arXiv:1003. 4615)
\bibitem{}ÊOwen, F. N., Hardee, P. E. \& Cornwell, T. J. 1989, ApJ, 340, 698
\bibitem{}ÊRossi, P. , Mignone, A,  Bodo, G, et al. 2008, A\&A , 488, 795
\bibitem{}ÊScheuer, P. A. G. \& Readhead, A. C., S. 1979, Nature, 277, 182 
\bibitem{}ÊSguera, V, Bassani, L.,  Malizia, A. et al.\ 2005, A\&A, 430, 107
\bibitem{}ÊSowards-Emmerd, D.,  Romani, R. W.,  Michelson, P.  F. 2003, ApJ 590, 109
\bibitem{}ÊSpinrad, H.,  Marr, J., Aguilar, L.,  Djorgovski, S.  1985  PASP, 97, 932
\bibitem{}ÊSreekumar, P., Bertsch, D.~L., Hartman, R.~C., Nolan, P.~L., \& Thompson, D.~J.\ 1999, Astroparticle Physics, 11, 221 
\bibitem{}ÊStawarz, {\L}., Ostrowski, M. 2002, ApJ, 578, 763
\bibitem{}ÊStawarz, {\L}., Sikora, M., \& Ostrowski, M.\ 2003, ApJ, 597, 186
\bibitem{}ÊStawarz, {\L}., Kneiske, T.~M., \& Kataoka, J.\ 2006, ApJ, 637, 693
\bibitem{}ÊSwain, M. R., Bridle, A. H., Baum, S. A.  1998,  ApJ, 507, L29
\bibitem{}ÊTaylor, G. B., Gugliucci, N. E., Fabian, A. C. et al.\ 2006 , MNRAS, 368, 1500
\bibitem{}ÊUrry C. M \& Padovani, P. 1995, PASP, 107, 803
\bibitem{}ÊWills, K. A., Morganti, R., Tadhunter, C. N. et al.  2004, MNRAS, 347, 771
\bibitem{}ÊWilkinson, P.N., Akujor, C. E., Cornwell, T.J., Saika, D. J. 1991, MNRAS, 248, 86
\end{thebibliography}
\end{document}